\documentclass[11pt]{article}
\usepackage{graphicx}
\usepackage{amsmath}
\usepackage{multirow}
\usepackage{amssymb}
\usepackage{amsthm}
\usepackage{booktabs}
\usepackage{bm}
\usepackage{natbib}
\usepackage{moreverb}
\usepackage{geometry}
\usepackage{bbm}

\usepackage{setspace}
\renewcommand{\P}{{\text{pr}}}
\newcommand{\E}{{{E}}}

\newcommand{\bH}{{{H}}}

\newcommand{\bx}{{{x}}}
\newcommand{\bX}{{X}}
\newcommand{\bZ}{{Z}}
\newcommand{\bz}{{z}}

\newcommand{\ba}{{{a}}}

\newcommand{\br}{{{r}}}

\newcommand{\bdelta}{{\delta}}

\newcommand{\brho}{{\varrho}}
\newcommand{\bu}{{{u}}}
\newcommand{\bb}{{{b}}}

\newcommand{\var}{{\text{var}}}

\newcommand{\bY}{{{Y}}}

\newcommand{\1}{1}

\renewcommand{\H}{{{H}}}

\newcommand{\bR}{{{R}}}
\newcommand{\bQ}{{{Q}}}
\newcommand{\boI}{{I}}

\newcommand{\cF}{{\mathcal{F}}}

\newcommand{\cZ}{{\mathcal{Z}}}

\newcommand{\bq}{{q}}

\newtheorem{theorem}{Theorem}
\newtheorem{proposition}{Proposition}
\newtheorem{corollary}{Corollary}

\newtheorem*{conditions*}{Sufficient Conditions}
\newtheorem{example}{Example}

\title{Testing weak nulls in matched observational studies}
\author{Colin B. Fogarty \thanks{Operations Research and Statistics Group, MIT Sloan School of Management, Massachusetts Institute of Technology, Cambridge MA 02142 (e-mail: \texttt{cfogarty@mit.edu})}}
\date{}

\begin{document}

\maketitle

\begin{abstract} 
We develop sensitivity analyses for weak nulls in matched observational studies while allowing unit-level treatment effects to vary. The methods may be applied to studies using any optimal without-replacement matching algorithm. In contrast to randomized experiments and to paired observational studies, we show for general matched designs that over a large class of test statistics, any valid sensitivity analysis for the entirety of the weak null must be unnecessarily conservative if Fisher's sharp null of no treatment effect for any individual also holds. We present a sensitivity analysis valid for the weak null, and illustrate why it is generally conservative if the sharp null holds through new connections to inverse probability weighted estimators. An alternative procedure is presented that is asymptotically sharp if treatment effects are constant, and that is valid for the weak null under additional restrictions which may be deemed benign by practitioners. Simulations demonstrate that this alternative procedure results in a valid sensitivity analysis for the weak null hypothesis under a host of reasonable data-generating processes. The procedures allow practitioners to assess robustness of estimated sample average treatment effects to hidden bias while allowing for unspecified effect heterogeneity in matched observational studies.
\end{abstract}
\section{Introduction}

Matching provides an appealing framework for inferring treatment effects in observational studies. Through the solution to an optimization problem, matching partitions individuals who have self-selected into treatment or control into matched sets on the basis of observed covariate information. In so doing one tries to create a fair comparison between treatment and control individuals, with the hope that after matching observed discrepancies in outcomes may be attributed to differences in treatment status rather than differences in confounding factors. Should the observational study be free of hidden bias, a successful application of matching would justify modes of inference valid for finely stratified experiments \citep{fog18mit}. Matching throws the fundamental differences between randomized experiments and observational studies into stark relief: in randomized experiments such modes of inference are valid by design, whereas in observational studies the same procedures require an assumption of no hidden bias that is both untestable and untenable.

The simplest and most prevalent form of matching is pair matching, where each matched set contains exactly one treated and one control individual who are similar on the basis of the observed covariates. Pair matching is but one of the matching algorithms available to practitioners, and is the least flexible of all matching algorithms. Alternatives include fixed ratio matching, variable ratio matching and full matching; see \citet{stu10, ros20} for a comprehensive overview of and worked examples using modern matching algorithms. These methods produce post-stratifications with a common structure: within each matched set, there is either one treated individual and many controls, or one control individual and many treated individuals, and each individual appears in at most one matched set. Of these methods, full matching makes use of the largest percentage of data while maintaining balance on observed covariates \citep{han04}. \citet{ros91} further illustrates that full matching enjoys desirable optimality properties in terms of minimizing aggregate covariate distance without discarding individuals. See \citet{stu08, kan16} for worked examples using full matching. In contrast, pair matching can be unnecessarily wasteful of data: if there are many more controls than treated individuals, a sizeable percentage of control individuals will be dropped from the analysis even if they are similar to treated units in the dataset. 

In a sensitivity analysis, the practitioner assesses the magnitude of hidden bias that would be required to overturn an observational study's finding of a treatment effect. Methods for sensitivity analysis in matched observational studies have traditionally focused on tests of sharp null hypotheses, i.e. hypotheses which impute the missing values for the potential outcomes.  The restrictiveness of this null hypothesis has been a point of contention in the analysis of randomized experiments and observational studies alike, with researchers in many domains instead desiring inference for Neyman's weak null that the average of the treatment effects equals zero \citep{ney35}. See \citet{sab14} and \citet{wu18} for more on arguments surrounding tests of sharp versus weak nulls.

The observed treated-minus-control difference in means has an expectation equal to the zero under both the sharp and weak null in a completely randomized design; however, the variance for the difference in means computed under the assumption of the sharp null may be too large or too small if instead only the weak null holds \citep{din17, loh17}. A single mode of inference that is both exact for the sharp null and asymptotically correct for the weak null is attained by instead employing a {studentized} randomization distribution. Rather than permuting the difference in means, one instead permutes the difference in means divided by a suitable standard error estimator; see \citet{bai19} for developments in adaptively paired experiments, and see \citet{jan97} and \citet{chu13, chu16} for related developments in robust permutation tests. Fundamentally, studentization only succeeds in unifying the modes of inference in randomized experiments because expectations of the treated-minus-control difference in means are the same under both null hypotheses. In an observational study, the difference in means would \textit{not} have expectation zero under either the sharp or weak null in the presence of hidden bias. When conducting a sensitivity analysis under a sharp null, one calculates the worst-case expectation for the test statistic being employed through the asymptotically separable algorithm of \citet{gas00}, reviewed in \S \ref{sec:cons}. This calculation makes explicit use of the sharp null hypothesis, and one may be concerned that the worst-case expectation when allowing for heterogeneous effects would be materially larger than that under the sharp null. Were this the case, studentization alone would be insufficient for aligning the methods for sensitivity analysis.

This work develops sensitivity analyses for flexible matched designs that are valid under effect heterogeneity. In contrast to the paired case analyzed in \citet{fog19student},  we show in Theorem \ref{thm:impossible} that over a large class of test statistics, no method of sensitivity analysis for general matched designs that tightly bounds the expectation assuming constant effects can also bound the worst-case expectation over the weak null. In matched observational studies, it is \textit{impossible} to unify the modes of inference for the weak null and the sharp null as can be done in randomized designs: any sensitivity analysis that is valid for the weak null must be unduly conservative for the sharp null, and any sensitivity analysis sharply bounding the expectation under constant effects must only be valid over a subset of the weak null. Theorem \ref{thm:impossible} frames the methodological developments of the paper, presenting two diverging paths to travel. The first path, explored in \S\ref{sec:Dtilde}, creates a sensitivity analysis that is always valid for the weak null but conservative for the sharp null.  The second, developed in \S\ref{sec:Dbar}, tightly bounds the expectation under constant effects and retains validity over interpretable restrictions on the weak null. We explore the fundamental source of divergence between the modes of inference through connections between matching estimators and inverse probability weighted estimators.  Inference is discussed in \S \ref{sec:inference}, while an improvement unique to binary outcomes is presented in \S\ref{sec:binary}, where it is shown that the worst-case expectation over the entirety of the weak null \textit{can} be efficiently computed. Simulation studies in \S \ref{sec:sim} indicate that the second procedure, valid over an interpretable subset of the weak null, continues to maintain validity for many reasonable data generating mechanisms, while a data example in \S \ref{sec:example} highlights the practical benefits of this method. All told, our developments enable practitioners using observational data to assess the robustness of their findings to hidden bias in any matched design even if effects are heterogeneous, providing methods for sensitivity analysis when practitioners are interested in average treatment effects.

\section{Notation for stratified experiments and observational studies}
\subsection{Notation for finely stratified designs}
\label{sec:notation}


There are $B$ independent matched sets formed on the basis of observed pretreatment covariates, the $i$th of which contains $n_i$ individuals. There are $N = \sum_{i=1}^Bn_i$ total individuals in the study. Each matched set contains one treated unit and $n_i-1$ control units. The developments in this article extend in a straightforward way to full matching; see problem 12 of \citet[ \S4]{obs} for further details.
  Let $Z_{ij}$ be an indicator of whether or not the $j$th individual in block $i$ received the treatment, such that $\sum_{i=1}^{n_i}Z_{ij}=1$ for all matched sets $i=1,...,B$. Along with a vector of measured covariates $\bx_{ij}$, each individual also has an unobserved covariate $0\leq u_{ij} \leq 1$. Let $r_{Tij}$ and $r_{Cij}$ denote the potential outcomes under treatment and control respectively for individual $ij$. The observed response is $R_{ij} = r_{Tij}Z_{ij} +r_{Cij}(1-Z_{ij})$, and the individual-level treatment effect $\tau_{ij} = r_{Tij} - r_{Cij}$ is not observable for any individual. Collect the potential outcomes, observed covariates and unobserved covariate for each individual into the set $\mathcal{F} = \{r_{Cij}, r_{Tij}, \bx_{ij}, u_{ij} : i=1,..,B; j=1,..,n_i\}$, the contents of which will be conditioned upon in the forthcoming developments as fixed properties of the observational study at hand. We will write $\bR = (R_{11},...,R_{Bn_B})^T$, to be the lexicographically-ordered vector containing all of the observed responses, $\bR_i = (R_{i1},...,R_{i{n_i}})^T$ for the observed responses in stratum $i$, and we will let the analogous notation hold for other vector quantities such as $u$ and $r_{Ci}$. 

\subsection{Treatment assignments in experiments and observational studies}
Let $\Omega = \{\bz: \sum_{i=1}^{n_i}z_{ij} =1, i=1,...,B\}$ be the set of $\prod_{i=1}^B n_i$ possible values of $\bZ = (Z_{11}, Z_{12},..., Z_{Bn_{B}})^T$ under a finely stratified design where one individual receives the treatment and $n_i-1$ receive the control, and let $\mathcal{Z}$ denote the event $\{\bZ \in \Omega\}$. In a finely stratified experiment \citep{fog18mit, pas19}, $\P(\bZ = \bz\mid\cF, \cZ) = \P(\bZ = \bz\mid\cZ)=  |\Omega|^{-1}$, and $\P(Z_{ij} = 1 \mid \cF, \cZ) = \P(Z_{ij} = 1 \mid \cZ)= 1/n_{i}$, where the notation $|A|$ denotes the cardinality of the set $A$. Before matching, individuals are assigned to treatment independently with unknown probabilities $\pi_{ij} = \P(Z_{ij}=1\mid \cF)$. While one may hope that $\pi_{ij}\approx \pi_{ij'}$ after matching, proceeding as such may produce misleading inference in due to the potential presence of unmeasured confounding. The model of \citet[][Chapter 4]{obs} states that individuals in the same matched set may differ in their odds of assignment to treatment by at most $\Gamma$, 
	\begin{equation}\label{eq:sensmodel}
	   \frac{1}{\Gamma} \leq \frac{\pi_{ij}(1 - \pi_{ij'})}{\pi_{ij'}(1 - \pi_{ij})} \leq \Gamma, (i=1,...,B; j,j' = 1,...,n_i).
	\end{equation} 
	
	The parameter $\Gamma$ controls the degree to which unmeasured confounding may have impacted the treatment assignment process. The value $\Gamma = 1$ returns a finely stratified experiment, while $\Gamma>1$ allows the unobserved covariates to influence the randomization distribution to a degree controlled by $\Gamma$. Returning attention to the matched structure by conditioning on $\cZ$, \citet{ros95} shows that (\ref{eq:sensmodel}) may be equivalently stated as \begin{equation*}
	    \P(\bZ = \bz \mid \cF,\, \cZ) = \frac{\exp\left(\gamma \bz^{T} \bu \right)}{\sum_{\bb \in \Omega}\exp\left(\gamma \bb^{T} \bu \right)} = \prod_{i=1}^B \frac{\exp\left(\gamma \sum_{j=1}^{n_i}z_{ij}u_{ij} \right)}{\sum_{j=1}^{n_i}\exp(\gamma u_{ij})},
	\end{equation*}
where $\gamma = \log(\Gamma)$ and $\bu$ lies in $\mathcal{U}$, the $N$-dimensional unit cube.

\subsection{The sample average treatment effect}
\label{sec:ATE}
The sample average treatment effect is the average of the individual-level treatment effects for the $N$ individuals in the study population, $\bar{\tau} = N^{-1}\sum_{i=1}^B\sum_{j=1}^{n_{i}}\tau_{ij} = \sum_{i=1}^B (n_i/N)\bar{\tau}_i$.  At $\Gamma=1$ the conventional unbiased estimator for $\bar{\tau}_i$, the average treatment effect for individuals in block $i$, is simply the observed difference in means between the treated and control individuals in block $i$,
$\hat{\tau}_i = \sum_{j=1}^{n_{i}}\left\{Z_{ij}r_{Tij} - (1-Z_{ij})r_{Cij}/(n_i-1)\right\}$. An unbiased estimator for the sample average treatment effect in a finely stratified experiment is $\hat{\tau} = \sum_{i=1}^B(n_i/N)\hat{\tau}_i$; however, this estimator may be substantially biased in the presence of unmeasured confounding. In what follows, it will be useful to define an additional quantity $\delta_{ij} = r_{Tij} - \sum_{j'\neq j}r_{Cij'}/(n_i-1)$ representing what the treated-minus-control difference in means would have been in stratum $i$ had individual $j$ received the treatment. We then have that $\bar{\tau}_i = \bar{\delta}_i$ and $\hat{\tau}_i = \sum_{j=1}^{n_i} Z_{ij}\delta_{ij}$.  
\section{Review: Asymptotic separability and sensitivity analysis assuming constant effects}\label{sec:cons}
Suppose interest lies in the null of a constant treatment effect at $\tau_0$, $\H_F^{(\tau_0)}: r_{Tij} = r_{Cij} + \tau_0\;\;\; (i=1,...,B; j=1,...,n_i),$
with a greater-than alternative. Consider using $\hat{\tau} - \tau_0$ as a test statistic. Under $H_F^{(\tau_0)}$, the adjusted responses $R_{ij} - Z_{ij}\tau_0$ equal $r_{Cij}$, imputing the missing values for the potential outcomes. Consequently, the values $\delta_{ij} - \tau_0 = R_{ij} - Z_{ij}\tau_0 - \sum_{j'\neq j}(R_{ij'} - Z_{ij'}\tau_0)/(n_i-1)$ are known under $H_F^{(\tau_0)}$ for all $ij$.

For a particular $\Gamma\geq 1$ and vector of unmeasured confounders $\bu$, the null distribution is
\begin{align}\label{eq:null}
&\P_{\bu}\{\hat{\tau} - \tau_0 \geq k\mid \cF, \cZ\} \nonumber\\&=  \sum_{\bz\in \Omega}\1\left\{\sum_{i=1}^B(n_i/N)\sum_{j=1}^{n_i}z_{ij}(\delta_{ij} -\tau_0)\geq k\right\}\prod_{i=1}^B \frac{\exp\left(\gamma \sum_{j=1}^{n_i}z_{ij}u_{ij} \right)}{\sum_{j=1}^{n_i}\exp(\gamma u_{ij})},
\end{align}   
where $\1(A)$ is an indicator that the condition $A$ is true. At $\Gamma=1$ (\ref{eq:null}) is simply the proportion of treatment assignments where the test statistic is greater than or equal to $k$, returning the usual randomization inference for $\hat{\tau}-\tau_0$ in a finely stratified experiment; this procedure is commonly referred to as the \textit{permutational $t$-test}. While the values $\delta_{ij}-\tau_0$ are known under $H_F^{(\tau_0)}$, for $\Gamma>1$ (\ref{eq:null}) remains unknown because of its dependence on the unmeasured covariates $\bu$. A sensitivity analysis using  $\hat{\tau} - \tau_0$ proceeds for a particular $\Gamma$ with the intent of finding the worst-case $p$-value over all possible $\bu$ based on the randomization distribution (\ref{eq:null}), $p^*_\Gamma(\tau_0) = \underset{\bu\in \mathcal{U}}{\max} \;\;\P_{\bu}\{\hat{\tau}-\tau_0 \geq \hat{\tau}^{obs}-\tau_0\mid \cF, \cZ\}$, where $\hat{\tau}^{obs}$ is the observed value of $\hat{\tau}$. The practitioner then increases $\Gamma$ until the test no longer rejects the null hypothesis, finding $\inf\{\Gamma: p^*_\Gamma(\tau_0) \geq \alpha\}$. This changepoint attests to the study's robustness to hidden bias.

Most test statistics for $H_F^{(\tau_0)}$ including $\hat{\tau} - \tau_0$ may be written in the form $\bZ^T\bq$ for some $\bq = \bq(\br_T-\tau_0, \br_C)$. Under $H_F^{(\tau_0)}$, $\bq(\br_T - \tau_0, \br_C)=$ $\bq(\bR - \bZ\tau_0, \bR - \bZ\tau_0)$, such that $\bq$ is determined by the observed data under the null. Rearrange the values $q_{ij}$ in each matched set such that $q_{i1} \leq q_{i2} \leq ... \leq q_{i_{n_i}}$. \citet{ros90} show that for a given $\Gamma$ in (\ref{eq:sensmodel}), the maximum value for $\P_{\bu}(\bZ^T\bq \geq k \mid \cF, \cZ)$ occurs at a vector $\bu$ of the form $u_{i1}=...=u_{i{a_i}} = 0$ and $u_{i{a_i}+1}=...=u_{i n_i} = 1$ for some $1 \leq a_i \leq n_{i}-1$ for each stratum. Let $\mathcal{U}^+_{i}$ denote this collection of $n_i-1$ binary vectors for the $i$th stratum. Unfortunately, this still results in $\prod_{i=1}^B(n_i-1)$ candidate values for the maximizer, rendering explicit calculation of $p^*_\Gamma(\tau_0)$ infeasible for general matched designs. \citet{gas00} show that under mild conditions an asymptotically valid upper bound to (\ref{eq:null}) can be attained by instead finding the vector $\bu_{i}$ in each matched set that maximizes the expectation for $\bZ_i^T\bq_i$. If multiple $\bu_i$ attain the same expectation, the one among these that maximizes the variance of $\bZ_i^T\bq_i$ is chosen.

Define $\mu_{\Gamma i} = \max_{\bu_i\in {\mathcal{U}}_i^+} {\sum_{j=1}^{n_i} \exp(\gamma u_{ij})q_{ij}}/\{{\sum_{j=1}^{n_i}\exp(\gamma u_{ij})}\},$ let $\tilde{\mathcal{U}}_{i}^+ \subseteq \mathcal{U}_i^+$ be the subset of binary vectors attaining the maximal expectation $\mu_{\Gamma i}$, and define \\$\nu_{\Gamma i} =  \max_{\bu_i\in \tilde{\mathcal{U}}_i^+} {\sum_{j=1}^{n_i} \exp(\gamma u_{ij})q_{ij}^2}/\{\sum_{j=1}^{n_i}\exp(\gamma u_{ij})\} - \mu_{\Gamma i}^2$. The asymptotic approximation to $\max_{\bu\in\mathcal{U}}\P_{\bu}(\bZ^T\bq \geq k \mid \cF, \cZ)$ returned by this procedure, called the separable algorithm, is
\begin{align}\label{eq:approx}
 1 - \Phi\left(\frac{\bZ^T\bq - \sum_{i=1}^B\mu_{\Gamma i}}{\sqrt{\sum_{i=1}^B\nu_{\Gamma i}}}\right),
\end{align} where $\Phi(\cdot)$ is the cumulative distribution function of the standard normal. Importantly, this asymptotic approximation reduces an optimization problem with $\prod_{i=1}^B(n_i-1)$ candidate solutions to $B$ tractable optimization problems which may be solved in isolation, each requiring enumeration of only $n_i-1$ candidate solutions.

The function $\bq(\br_T - \tau_0, \br_C)$ giving rise to $\hat{\tau} - \tau_0$ is $\bq(\br_T-\tau_0, \br_C)_{ij} = (n_i/N)(\delta_{ij} -\tau_0)= (n_i/N)\left\{r_{Tij} - \tau_0 - \sum_{j'\neq j}r_{Cij'}/(n_i-1)\right\}$. Let $\tilde{p}_\Gamma(\tau_0)$ be the probability in (\ref{eq:approx}) when the separable algorithm is applied to this choice of $\bq$ at $\Gamma$, such that $\tilde{p}_\Gamma(\tau_0)$ provides an asymptotic approximation to the true worst-case $p$-value $p^*_\Gamma(\tau_0)$ under the sharp null $H_F^{(\tau_0)}$. The resulting large-sample sensitivity analysis using the difference in means is
 $\varphi_{DiM}^{(\tau_0)}(\alpha,\Gamma) = \1\{\tilde{p}_\Gamma(\tau_0) \leq \alpha\},$ and \citet{gas00} show that  $\limsup_{B\rightarrow\infty}E\{\varphi_{DiM}^{(\tau_0)}(\alpha,\Gamma)\mid \cF, \cZ\}\leq \alpha$ under mild conditions if (\ref{eq:sensmodel}) holds at $\Gamma$ and $H_F^{(\tau_0)}$ is true.

\section{Challenges facing sensitivity analysis under effect heterogeneity}\label{sec:challenge}

\subsection{The inadequacy of studentization in general matched designs}\label{sec:inadequate}
Suppose interest instead lies in the null hypothesis that the sample average of the treatment effects equals some value $\tau_0$, $H_N^{(\tau_0)}: \bar{\tau} = \tau_0$, while leaving the individual treatment effects $\tau_{ij}$ $(i=1,...,B; j=1,...,n_i)$ otherwise unspecified. $H_N^{(0)}$ reflects Neyman's weak null hypothesis of no treatment effect on average for the individuals in the study \citep{ney35, din17}. $H_N^{(\tau_0)}$ is a composite null hypothesis, and $H_F^{(\tau_0)}$ is one particular element. We first consider the properties of the permutational $t$ test described in \S\ref{sec:cons} when $H_N^{(\tau_0)}$ holds but $H_F^{(\tau_0)}$ does not in observational studies $(\Gamma>1)$. After doing so, we move beyond the permutational $t$ statistic and present a general theorem regarding the use of sensitivity analyses conducted by way of the separable algorithm when effects are not actually constant.

We now demonstrate that the permutational $t$-based sensitivity analysis need not bound the worst-case expectation, rendering studentization, the conventional fix for randomized experiments, insufficient to align the modes of inference.  Define $\varrho_{ij} = \P(Z_{ij}=1\mid \cF, \cZ) = \exp(\gamma u_{ij})/\{\sum_{j'=1}^{n_i}\exp(\gamma u_{ij'})\}$  as the conditional probability of assignment to treatment for the $ij$th individual when (\ref{eq:sensmodel}) holds at $\Gamma$. Under $H_N^{(\tau_0)}$ $R_{ij} - Z_{ij}\tau_0$ does not generally equal $r_{Cij}$ when $Z_{ij}=1$. Instead, it equals $r_{Cij} + \tau_{ij} - \tau_0$. The true value of $\delta_{ij} - \tau_0$ is only known for the index $k$ such that $Z_{ik}=1$ in the observational study at hand. If $Z_{ik}=1$, the separable algorithm in \S \ref{sec:cons} proceeding under the assumption of constant effects generates a candidate worst-case expectation in stratum $i$ using improperly imputed values for $\delta^{(k)}_{ij}$ of the form
\begin{align*}
\delta^{(k)}_{ij} - \tau_0&= \begin{cases} \delta_{ij} - \tau_0 & j = k\\ \delta_{ij} - \tau_{ij} -  (\tau_{ik} - \tau_0)/(n_i-1) & \text{otherwise},\end{cases}
\end{align*} 
such that $\delta^{(k)}_{ij} - \tau_0 \neq \delta_{ij} - \tau_0$ for $j\neq k$ under $H_N^{(\tau_0)}$. For each $k=1,...,n_i$, let $\brho^{(k)}_{i}$ be the probabilities giving rise to the worst-case expectation when the separable algorithm is applied with $Z_{ik} = 1$, and let ${\vartheta}_{\Gamma ik} = \sum_{j=1}^{n_i}\varrho^{(k)}_{ij}\delta^{(k)}_{ij}$ be the worst-case expectation returned by the separable algorithm assuming that treatment effects are constant at $\tau_0$ when $Z_{ik}=1$. The random variable stemming from the asymptotically separable algorithm in the $i$th set is $T^{(\tau_0)}_{\Gamma i} = \sum_{j=1}^{n_i}Z_{ij}(\delta_{ij} - \tau_0 - {\vartheta}_{\Gamma ij})$, and let $\bar{T}^{(\tau_0)}_\Gamma = \sum_{i=1}^B(n_i/N)T^{(\tau_0)}_{\Gamma i}$ be their weighted average. Let $\bu$ and $\brho$ be the true vectors of unmeasured confounders and the corresponding conditional probabilities. For the separable algorithm to furnish an asymptotically valid sensitivity analysis under effect heterogeneity, it is necessary that in the limit $E_{\bu}(\bar{T}^{(\tau_0)}_{\Gamma} \mid \cF, \cZ) \leq 0$ when (\ref{eq:sensmodel}) holds at $\Gamma$. Unfortunately, this need not be the case.
\begin{proposition}\label{prop:Tbar}
Suppose (\ref{eq:sensmodel}) holds at $\Gamma$ and that $H_N^{(\tau_0)}$ is true. Then, 
\begin{align*}
E_\bu(\bar{T}^{(\tau_0)}_{\Gamma}\mid \cF, \cZ) &\leq \sum_{i=1}^B\frac{n_i^2}{N(n_i-1)}\sum_{j=1}^{n_i}\varrho_{ij}(1-\varrho_{ij})(\tau_{ij}-\tau_0).
\end{align*}
Moreover, there exist allocations of potential outcomes satisfying $H_N^{(\tau_0)}$ with unmeasured confounders $\bu$ such that $E_\bu(\bar{T}^{(\tau_0)}_{\Gamma}\mid \cF, \cZ) > 0$ when (\ref{eq:sensmodel}) holds at $\Gamma$.
\end{proposition} Proposition \ref{prop:Tbar} is proved in the appendix. At $\Gamma=1$, $\varrho_{ij}(1-\varrho_{ij}) = (n_i-1)/n_i^2$, returning the usual result that $E(\hat{\tau}_i - \tau_0\mid \cF, \cZ) = \bar{\tau}_i - \tau_0$ in a finely stratified experiment. For a paired observational study with $n_i=2$, at the worst-case vector of assignment probabilities $\brho^*$ we have that $\varrho^*_{ij}(1-\varrho^*_{ij}) = \Gamma/(1+\Gamma)^2$ for all $ij$. As a result, the worst-case expectation is bounded above by zero under $H_N^{(\tau_0)}$ as proved in Theorem 1 of \citet{fog19student}. However, for $\Gamma > 1$ and $n_i > 2$, the unit-level treatment effects may be unequally weighted by the conditional selection probabilities even under the worst-case vector of unmeasured confounding. Should these selection probabilities correlate with the individual-level treatment effects in an adverse manner, the expectation may fail to be controlled. See the appendix for a worked example where this behavior occurs. As the expectation is not properly controlled by the separable algorithm, studentizing $\bar{T}^{(\tau_0)}_\Gamma$ will not restore asymptotic validity to the resulting procedure.


\subsection{An impossibility result for general matched designs}
The previous section illustrates for matched designs beyond pair matching, a sensitivity analysis using $\hat{\tau} - \tau_0$ as a test statistic under the assumption of constant effects need not yield a valid sensitivity analysis if effects are instead heterogeneous. While the permutational $t$ test cannot be employed to this end, one may hope that another test statistic could achieve this objective. As we now show, this is regrettably not the case. Over a large class of test statistics that are functions of the stratum-wise treated-minus-control mean differences $\hat{\tau}_i$, a sensitivity analysis assuming constant effects cannot be relied upon to bound the worst-case expectation if the treatment effects are instead heterogeneous.

For $k$ scalar valued, let $h_{\Gamma n_i}(k)$ be a monotone nondecreasing, non-constant function allowed to depend upon both the stratum size $n_i$ and the value of $\Gamma$ at which the sensitivity analysis is being conducted. In the $i$th stratum, consider the statistic $h_{\Gamma n_i}(\hat{\tau}_i - \tau_0)$. Let $\mu_{ \Gamma i} = \mu_{\Gamma i}(h_{\Gamma n_i}, \bR_{i} - \bZ_i\tau_0)$ be the worst-case expectation for $h_{\Gamma n_i}(\hat{\tau}_i - \tau_0)$ returned by the separable algorithm at $\Gamma$ under $H_F^{(\tau_0)}$. If (\ref{eq:sensmodel}) holds at $\Gamma$, by construction we have that $B^{-1}\sum_{i=1}^BE\left\{h_{\Gamma n_i}(\hat{\tau}_i-\tau_0) - \mu_{\Gamma i}\mid \cF, \cZ \right\} \leq 0$ when $H_F^{(\tau_0)}$ actually holds. If $H_F^{(\tau_0)}$ is instead false but $H_N^{(\tau_0)}$ is true, $\mu_{\Gamma i}$ varies over $\bz \in \Omega$ as $\bR_{i} - \bZ_i\tau_0 \neq \br_{Ci}$. The sensitivity analysis assuming constant effects would also bound the worst-case expectation under $H_N^{(\tau_0)}$ for a given collection of functions $\{h_{\Gamma n_i}\}_{n_i\geq 2}$ if $B^{-1}\sum_{i=1}^BE\left\{h_{\Gamma n_i}(\hat{\tau}_i-\tau_0) - \mu_{\Gamma i}\mid \cF, \cZ\right\} \leq 0$ under $H_N^{(\tau_0)}$ when (\ref{eq:sensmodel}) holds at $\Gamma$ for any $\Gamma$. The following result shows that this cannot be guaranteed over all elements of the composite null $H_N^{(\tau_0)}$ for general matched designs.

\begin{theorem}\label{thm:impossible}
For any collection of nondecreasing, nonconstant functions $\{h_{\Gamma n_i}\}_{n_i\geq 2}$, there exist combinations of stratum sizes $n_i$ $(i=1,...,B)$, degrees of hidden bias $\Gamma$, and values for potential outcomes satisfying $H_N^{(\tau_0)}$ such that if (\ref{eq:sensmodel}) holds at $\Gamma$,
\begin{align*}\max_{\bu \in \mathcal{U}}\;\; \frac{1}{B}\sum_{i=1}^BE_{\bu_i}\left\{h_{\Gamma n_i}(\hat{\tau}_i-\tau_0) - \mu_{\Gamma i}(h_{\Gamma n_i}, \bR_{i} - \bZ_i\tau_0)\mid \cF, \cZ\right\} > 0.\end{align*}
That is, no sensitivity analysis using the worst-case expectation under $H_F^{(\tau_0)}$ can bound the worst-case expectation under $H_N^{(\tau_0)}$ when effects are heterogeneous for all possible combinations of stratum sizes, degrees of hidden bias, and for all elements of the weak null.
\end{theorem}
The proof is presented in the appendix. It is constructive, creating for any $n_i \geq 3$ an allocation of potential outcomes satisfying $H_N^{(\tau_0)}$, but where for any non-decreasing, non-constant function $h_{\Gamma n_i}(\hat{\tau}_i - \tau_0)$ there exists a $\Gamma$ such that the worst-case expectation under $H_F^{(\tau_0)}$ fails to bound the worst-case expectation under $H_N^{(\tau_0)}$. The theorem shows that for a large class of test statistics, including any weighted average of treated-minus-control differences in means across strata, it is in general impossible to devise an upper bound for $B^{-1}\sum_{i=1}^B E_{\bu_i}\{h_{\Gamma n_i}(\hat{\tau}_i - \tau_0)\mid \cF, \cZ\}$ that is simultaneously tight under both the sharp null and the weak null outside the confines of pair matching. If the bound is tight when effects are assumed constant at $\tau_0$, it must generally not bound the expectation when instead only $\bar{\tau} = \tau_0$. If the bound is tight when testing $\bar{\tau} = \tau_0$ while accommodating effect heterogeneity, it must be a strict upper bound when effects are actually constant at $\tau_0$. Unlike in randomized experiments, the modes of sensitivity analysis for constant effects and averages of heterogeneous effects cannot be unified without the procedure being unnecessarily conservative under the sharp null. 

\subsection{Towards sensitivity analysis for weak nulls}\label{sec:flow}

Suppose that (\ref{eq:sensmodel}) holds at $\Gamma$ and that $\bar{\tau} = \tau_0$, such that $H_N^{(\tau_0)}$ is true. Let $L^{(\tau_0)}_\Gamma = B^{-1}\sum_{i=1}^BL^{(\tau_0)}_{\Gamma i}$ be any candidate test statistic, and let $\bu$ be the true, but unknowable, vector of unmeasured confounders. Under suitable regularity conditions on $\cF$ a central limit theorem would apply to $L^{(\tau_0)}_\Gamma$, such that 
\begin{align}\label{eq:idealprob}
\underset{B\rightarrow\infty}{\lim} \P\left(\frac{L^{(\tau_0)}_\Gamma - E_{\bu}(L^{(\tau_0)}_\Gamma  \mid \cF, \cZ)}{\var_{\bu}(L^{(\tau_0)}_\Gamma  \mid \cF, \cZ)^{1/2}} \geq k\right) &= 1-\Phi(k)
\end{align} for any $k$. The deviate within (\ref{eq:idealprob}) cannot be used in practice, as the expectation and variance both depend upon the unmeasured confounders and the unknown potential outcomes. A straightforward observation provides that
\begin{align}\label{eq:bound1}
\frac{L^{(\tau_0)}_\Gamma - E_{\bu}(L^{(\tau_0)}_\Gamma  \mid \cF, \cZ)}{\var_{\bu}(L^{(\tau_0)}_\Gamma  \mid \cF, \cZ)^{1/2}} &\geq \frac{L^{(\tau_0)}_\Gamma -\max_{\bu\in\mathcal{U}} E_{\bu}(L^{(\tau_0)}_\Gamma  \mid \cF, \cZ)}{\var_{\bu}(L^{(\tau_0)}_\Gamma  \mid \cF, \cZ)^{1/2}}.
\end{align}The right-hand side of (\ref{eq:bound1}) simply replaces the true but unknowable expectation with its worst-case value at a given $\Gamma$. When testing the sharp null $H_F^{(\tau_0)}$, the separable algorithm of \citet{gas00} outlined in \S \ref{sec:cons} can be used to a construct this worst-case expectation. Unfortunately, this explicit construction requires assuming a sharp null so that the unknown potential outcomes are imputed. When testing a weak null hypothesis, the missing potential outcomes remain unspecified, such that the worst-case expectation cannot be calculated. Nonetheless, we will show that while it cannot be explicitly calculated, with a smart choice of test statistic the worst-case expectation $\max_{\bu\in\mathcal{U}} E_{\bu}(L^{(\tau_0)}_\Gamma  \mid \cF, \cZ)$ may itself be bounded over the weak null $H_N^{(\tau_0)}$ in a way that facilitates an implementable sensitivity analysis. In \S \ref{sec:Dtilde}, we present a test statistic whose worst-case expectation is tightly bounded above by zero under the weak null. We then illustrate precisely why this bound is conservative if effects are instead constant.  In \S \ref{sec:Dbar}, we present an alternative test statistic whose expectation is tightly bounded above by zero under the sharp null. Further, the worst-case expectation continues to be bounded above by zero over a subset of the weak null which may be viewed as benign by practitioners.   

Even if we knew the worst-case expectation in the right-hand side of (\ref{eq:bound1}), the deviate could not be deployed in practice due to the unknown variance. We show in \S \ref{sec:SE} that the observed data can be used to construct a conservative variance estimator, despite the true variance depending on both the missing potential outcomes and the unmeasured confounders. That is, for the statistics in \S\S\ref{sec:Dtilde}-\ref{sec:Dbar} we construct a sample-based standard error $\text{se}(L^{(\tau_0)}_\Gamma)$ such that $\E_\bu\{\text{se}^2(L^{(\tau_0)}_\Gamma)\mid \cF, \cZ\} \geq \var_{\bu}(L^{(\tau_0)}_\Gamma  \mid \cF, \cZ)$, and such that the squared standard error, suitably scaled, converges in probability to its expectation under mild conditions. That is, for $L^{(\tau_0)}_\Gamma > 0$ and using the squared standard error's expectation,
\begin{align}\label{eq:bound2}
\frac{L^{(\tau_0)}_\Gamma - E_{\bu}(L^{(\tau_0)}_\Gamma  \mid \cF, \cZ)}{\var_{\bu}(L^{(\tau_0)}_\Gamma  \mid \cF, \cZ)^{1/2}} &\geq \frac{L^{(\tau_0)}_\Gamma -\max_{\bu\in\mathcal{U}} E_{\bu}(L^{(\tau_0)}_\Gamma  \mid \cF, \cZ)}{\E_\bu\{\text{se}^2(L^{(\tau_0)}_\Gamma)\mid \cF, \cZ\}^{1/2}}.
\end{align} A deviate replacing the worst-case expectation in the right-hand side of (\ref{eq:bound2}) with a computable upper bound and replacing $\E_\bu\{\text{se}^2(L^{(\tau_0)}_\Gamma)\mid \cF, \cZ\}^{1/2}$ with the estimator $\text{se}(L^{(\tau_0)}_\Gamma)$ would have its upper tail probability bounded above by a standard normal in the limit, in turn facilitating an asymptotically valid sensitivity analysis for $H_N^{(\tau_0)}$.

\section{A valid sensitivity analysis with heterogeneous effects}\label{sec:Dtilde}
\subsection{Interval restrictions on the stratum-wise assignment probabilities}\label{sec:interval}
For the time being, consider a further restriction on the allowed assignment probabilities in (\ref{eq:sensmodel}). Imagine that for each stratum $i$ and for any $\Gamma$, there exist known constants $\kappa_{\Gamma i}$ such that, in addition to the constraint $\sum_{j=1}^{n_i}\varrho_{ij} = 1$ for all $i$,
\begin{align}\label{eq:interval}
\kappa^{-1}_{\Gamma i} \leq \varrho_{ij} \leq \Gamma \kappa^{-1}_{\Gamma i}.
\end{align}The restriction (\ref{eq:interval}) holding at $\Gamma$ for prespecified constants $\kappa_{\Gamma i}$ implies that Rosenbaum's model (\ref{eq:sensmodel}) also holds at $\Gamma$. Unfortunately, the converse is not true: while there will be a value $\kappa_{\Gamma i}$ such that (\ref{eq:interval}) holds, its particular value is not specified when assuming that (\ref{eq:sensmodel}) holds at $\Gamma$. See \citet{aro12} and \citet{mir17} for a related model for inference with unknown probabilities of sample selection.

Suppose interest lies in the null hypothesis $H_N^{(\tau_0)}$ with (\ref{eq:interval}) holding at $\Gamma$, and consider 
\begin{align}\label{eq:D}
{D}^{(\tau_0)}_{\Gamma i} = \hat{\tau}_i - \tau_0 - \left(\frac{\Gamma-1}{1+\Gamma}\right)|\hat{\tau}_i-\tau_0|.
\end{align}
${D}^{(\tau_0)}_{\Gamma i}$ is precisely the treated-minus-control mean difference in stratum $i$, subtracted by the worst-case expectation under a paired design for $H_F^{(\tau_0)}$ when (\ref{eq:sensmodel}) holds at $\Gamma$. \citet{fog19student} demonstrated that ${D}^{(\tau_0)}_{\Gamma i}$ can be used to construct an asymptotically valid sensitivity analysis for  $H_N^{(\tau_0)}$ in paired observational studies. Here we show the importance of this random variable in general matched designs with heterogeneous effects. 

Consider the $\kappa_\Gamma$-weighted average 
\begin{align} K^{(\tau_0)}_\Gamma  &= N^{-1}\sum_{i=1}^B\kappa_{\Gamma i}{D}^{(\tau_0)}_{\Gamma i}\label{eq:K}\end{align}.\begin{proposition}\label{prop:interval}
Suppose (\ref{eq:interval}) holds at $\Gamma$ for known values $\kappa_{\Gamma i}$ $(i=1,...,B)$. Then, 
\begin{align*} E({D}^{(\tau_0)}_{\Gamma i}\mid \cF, \cZ) &\leq \left(\frac{n_i}{\kappa_{\Gamma i}}\right)\left(\frac{2\Gamma}{1+\Gamma}\right)(\bar{\tau}_i - \tau_0);\\
E(K^{(\tau_0)}_{\Gamma}\mid \cF, \cZ) &\leq \left(\frac{2\Gamma}{1+\Gamma}\right)(\bar{\tau} - \tau_0)
\end{align*}\end{proposition} If (\ref{eq:interval}) holds at $\Gamma$ for known constants $\kappa_{\Gamma i}$, then $E(K^{(\tau_0)}_{\Gamma i}\mid \cF, \cZ)\leq 0$ under $H_N^{(\tau_0)}$. Test statistics of the form $K^{(\tau_0)}_\Gamma$ can control the worst-case expectation with heterogeneous effects under the interval restriction (\ref{eq:interval}).

\subsection{A connection with inverse probability weighted estimators}\label{sec:inverse probability weighted}
We now draw connections between the random variable ${D}_{\Gamma i}^{(\tau_0)}$ and an inverse probability weighted estimator under the restriction (\ref{eq:interval}). Consider the random variable
\begin{align}\label{eq:IPW}
W^{(\tau_0)}_{\Gamma i} &= \frac{1}{n_i}\left(\underset{\kappa_{\Gamma i}^{-1}\leq p_i \leq \Gamma\kappa^{-1}_{\Gamma i}}{\min}\;\; \frac{\hat{\tau}_i - \tau_0}{p_i}\right),
\end{align}
which weights $\hat{\tau}_i-\tau_0$ in the worst possible way for inference with a greater-than alternative under the interval restriction (\ref{eq:interval}). If we had access to the true stratum-wise treatment assignment probabilities $\brho_{i}$, by classical results on inverse probability weighted estimators \citep{hor52} an unbiased estimator of $\bar{\tau}_i - \bar{\tau}_0$ would be 
\begin{align*}
IPW^{(\tau_0)}_i &= \frac{1}{n_i}\left(\frac{\hat{\tau}_i - \tau_0}{\sum_{j=1}^{n_i}Z_{ij}\varrho_{ij}}\right) = \frac{1}{n_i}\left(\frac{\sum_{j=1}^{n_i}Z_{ij}(\delta_{ij} - \tau_0)}{\sum_{j=1}^{n_i}Z_{ij}\varrho_{ij}}\right).
\end{align*} As we do not know $\brho_i$ due to its dependence on the unmeasured confounders, we cannot employ $IPW^{(\tau_0)}_i$ for any $i$. The statistic $W^{(\tau_0)}_{\Gamma i}$ instead weights $(\hat{\tau}_i-\bar{\tau}_0)$ by the worst-case conditional probability under (\ref{eq:interval}), weighting by the largest probability if $\hat{\tau}_i - \tau_0 \geq 0$ and by the smallest probability when $\hat{\tau}_i - \tau_0 < 0$, resulting in an expectation no larger than $\bar{\tau}_i - \tau_0$. 

\begin{proposition}\label{prop:equiv}
\begin{align*} W^{(\tau_0)}_{\Gamma i}  &= \left(\frac{\kappa_{\Gamma i}}{n_i}\right)\left(\frac{1+\Gamma}{2\Gamma}\right){D}^{(\tau_0)}_{\Gamma i},\end{align*} such that ${D}^{(\tau_0)}_{\Gamma i}$ is proportional to the worst-case $IPW$ estimator given $\kappa^{-1}_{\Gamma i} \leq \varrho_{ij}\leq \Gamma \kappa^{-1}_{\Gamma i}$.\end{proposition} This connection to an inverse probability weighted estimator provides intuition for why ${D}^{(\tau_0)}_{\Gamma i}$ can be used to control the worst-case expectation in arbitrary matched designs, despite its functional form having been motivated by sensitivity analyses in paired designs. 

\subsection{A valid sensitivity analysis for the weak null under Rosenbaum's model}\label{sec:valid}
Suppose instead one wants to perform a sensitivity analysis for the sample average treatment effect without imposing a particular interval restriction as in (\ref{eq:interval}). That is, one simply desires a sensitivity analysis for the sample average treatment effect under the assumption that (\ref{eq:sensmodel}) holds at $\Gamma$. For a given stratum size $n_i$, under (\ref{eq:sensmodel}) the resulting conditional assignment probabilities are confined to 
\begin{align}\label{eq:intervalrosorig} \frac{1}{\Gamma(n_i-1) + 1} \leq \varrho_{ij} \leq \frac{\Gamma}{(n_i-1)+\Gamma},\end{align} or in an equivalent form akin to (\ref{eq:interval}),
\begin{align}\label{eq:intervalros} 
\tilde{\kappa}_{\Gamma n_i}^{-1} \leq \varrho_{ij} \leq \Gamma_{n_i}\tilde{\kappa}_{\Gamma n_i}^{-1};\;\;\;\tilde{\kappa}_{\Gamma n_i} = \Gamma(n_i -1) + 1;\;\; \Gamma_{n_{i}} = \Gamma \left\{\frac{\Gamma(n_i-1) + 1}{(n_i-1)+\Gamma}\right\}.
\end{align} For matched pairs, we have that $\Gamma_{n_{i}} = \Gamma$; however, for $n_i>2$, $\Gamma_{n_{i}} > \Gamma$ if $\Gamma>1$, and in fact $\Gamma_{n_{i}}\rightarrow \Gamma^2$ as $n_i\rightarrow\infty$. Consider the random variable
\begin{align}\label{eq:Dtilde}
\tilde{D}^{(\tau_0)}_{\Gamma i} &= \hat{\tau} - \tau_0 - \left(\frac{\Gamma_{n_i} - 1}{1+\Gamma_{n_{i}}}\right)|\hat{\tau} - \tau_0|,
\end{align}
and define the $\tilde{\kappa}_\Gamma$-weighted average $\tilde{K}^{(\tau_0)}_{\Gamma} = N^{-1}\sum_{i=1}^B\tilde{\kappa}_{\Gamma n_i}\tilde{D}^{(\tau_0)}_{\Gamma i}$.
\begin{theorem}\label{thm:weak}
Suppose (\ref{eq:sensmodel}) holds at $\Gamma$ and that $H_N^{(\tau_0)}$ is true. Then $E_\bu(\tilde{K}^{(\tau_0)}_{\Gamma} \mid \cF, \cZ)\leq 0,$ such that the worst-case expectation is controlled through this test statistic. Further, there exist potential outcomes within $H_N^{(\tau_0)}$ and unmeasured confounders $\bu$ such that $E_\bu(\tilde{K}^{(\tau_0)}_{\Gamma} \mid \cF, \cZ) = 0$.
\end{theorem}
Theorem \ref{thm:weak}, proved in the appendix, demonstrates that the random variable $\tilde{K}^{(\tau_0)}_{\Gamma}$ has a known upper bound on its expectation when the sample average treatment effect equals $\tau_0$ and (\ref{eq:sensmodel}) holds at $\Gamma$, even when allowing for heterogeneous effects. Further, it states that the bound is tight for certain elements of $H_N^{(\tau_0)}$, such that it cannot be improved upon without further restrictions on $H_N^{(\tau_0)}$ or on the unmeasured confounders $\bu$. The example used to prove Theorem \ref{thm:impossible} represents an instance where the worst-case expectation equals zero.

\subsection{Explaining the divergence for sharp and weak nulls: Incompatibility}\label{sec:diverge}
Theorem \ref{thm:impossible} implies a necessary divergence between sensitivity analyses assuming constant effects and those allowing for heterogeneous effects. In particular, we know that because $\tilde{K}^{(\tau_0)}_{\Gamma}$ has an expectation bounded above by zero under $H_N^{(\tau_0)}$ when (\ref{eq:sensmodel}) holds at $\Gamma$, it is necessary that $E(\tilde{K}^{(\tau_0)}_{\Gamma}\mid \cF, \cZ) < 0$ under $H_F^{(\tau_0)}$ when (\ref{eq:sensmodel}) holds at $\Gamma$ unless we have a paired observational study. The following proposition helps explain why this occurs.

\begin{proposition}\label{prop:zero}
Consider $A_i = \sum_{j=1}^{n_i}Z_{ij}\left\{q_{ij} - |q_{ij}|(\Gamma-1)/(1+\Gamma)\right\},$ where $q_{ij}$ are any constants such that $\sum_{j=1}^{n_i}q_{ij} = 0$. Suppose (\ref{eq:sensmodel}) holds at $\Gamma$. Then,
$E_\bu(A_i\mid \cF, \cZ) \leq 0$, and $E_\bu(A_i\mid \cF, \cZ) = 0$ if $u_{ij} = \1(q_{ij} \geq 0)$.
\end{proposition}

Under $H_F^{(\tau_0)}$, we have $\sum_{i=1}^{n_i}(\delta_{ij} - \tau_0) = 0$ as $\bar{\delta}_i = {\tau}_0$. Therefore, under constant effects, Proposition \ref{prop:zero} along with the form for $\tilde{D}^{(\tau_0)}_{\Gamma i}$ (\ref{eq:Dtilde}) imply that $E(\tilde{D}^{(\tau_0)}_{\Gamma i}\mid \cF, \cZ) \leq 0$ even when (\ref{eq:sensmodel}) holds at $\Gamma_{n_i} = \Gamma \{\Gamma(n_i-1) + 1\}/\{(n_i-1)+\Gamma\}$, which is strictly greater than $\Gamma$ for $n_i > 2$ and $\Gamma>1$. Proceeding as though the worst-case expectation for $\tilde{K}^{(\tau_0)}_{\Gamma}$ is zero would thus be unnecessarily conservative under $H_F^{(\tau_0)}$ when conducting a sensitivity analysis at $\Gamma$, and one could find a less conservative bound by applying the separable algorithm in \S \ref{sec:cons}.

Interpreting $\tilde{D}^{(\tau_0)}_{\Gamma i}$ as a worst-case inverse probability weighted estimator as in (\ref{eq:IPW}) is helpful in understanding both the source of the conservativeness when assuming constant effects, and the reason it cannot be overcome when effects are heterogeneous. For each $ij$, we know that value $\varrho_{ij}$ minimizing $(\delta_{ij}-\tau_0)/\varrho_{ij}$ subject to (\ref{eq:intervalrosorig}) is $\tilde{\varrho}_{ij} = \Gamma/(n_i-1 +\Gamma)$ if $(\delta_{ij}-\tau_0) > 0$;  $\tilde{\varrho}_{ij} = 1/(\Gamma(n_i-1) +1 )$ if $(\delta_{ij}-\tau_0) < 0$; and is any feasible $\tilde{\varrho}_{ij}$ if $\delta_{ij} - \tau_0 = 0$. Observe that $\sum_{j=1}^{n_i}\tilde{\varrho}_{ij}$ need not equal 1. That is, the vector ${\tilde{\varrho}}_i$ used when weighting need not be a valid probability distribution for $\bZ_i\mid \cF, \cZ$, and it will not be when $n_i > 2$ so long as $\delta_{ij}$ is not constant at $\tau_0$ for all $j=1,...,n_i$. The following numerical provides an illustration of this.
\begin{example}
Suppose $n_i=3$, and that the potential values for $\hat{\tau}_i$ given $\cF, \cZ$ are $\delta_{i1} = 5, \delta_{i2}=-2, \delta_{i3}=-3$ such that $\bar{\tau}_i=0$, and we conduct inference at $\Gamma=2$. The constraints in (\ref{eq:intervalrosorig}) at $\Gamma=2$ and $n_i=3$ imply that $1/5\leq \varrho_{ij}\leq 1/2$, and the equivalent worst-case IPW estimator $\tilde{W}^{(0)}_{2i}$ from (\ref{eq:IPW}) for $\tilde{D}^{(0)}_{2 i}$ takes the form
\begin{align*}
\tilde{W}^{(0)}_{2i} &= \frac{1}{3}\left(\frac{\hat{\tau}_i}{\left(\frac{1}{2}\right)\1(\hat{\tau}_i \geq 0) + \left(\frac{1}{5}\right)\1(\hat{\tau}_i < 0)}\right).
\end{align*}
The mapping from potential values of $\hat{\tau}_i$ to corresponding worst-case ``probabilities'' is $\delta_{i1} = 5 \Rightarrow \tilde{\varrho}_{i1} = 1/2$; $\delta_{i2} = -2 \Rightarrow \tilde{\varrho}_{i2} = 1/5$; $\delta_{i3}= -3 \Rightarrow \tilde{\varrho}_{i3} = 1/5$. Observe that $\sum_{j=1}^{3}\tilde{\varrho}_{ij} = 9/10$, which is not a probability distribution. If (\ref{eq:sensmodel}) truly holds at $\Gamma=2$, we then have that $E(\tilde{W}^{(0)}_{2i}\mid \cF, \cZ) < 0$. 
\end{example}

When assuming the sharp null $H_F^{(\tau_0)}$ all possible values of $\delta_{ij} - \tau_0$ are known from the observed data, such that the separable algorithm can be used to produce a worst-case expectation based upon a valid probability distribution, imposing the constraint $\sum_{j=1}^{n_i}\varrho_{ij}^*=1$ for all $i$. Under $H_N^{(\tau_0)}$, only the $\delta_{ij}-\tau_0$ for which $Z_{ij}=1$ in the observed study is known for each $i$, leaving $n_i-1$ values for $\delta_{ij}-\tau_0$ unknown in each matched set. Despite knowing that $\tilde{\brho}_i$ will not correspond to a valid probability distribution, one cannot act upon this knowledge when conducting inference for $H_N^{(\tau_0)}$ without risking an anti-conservative procedure for the data set at hand. This issue is referred to as \textit{incompatibility} and occurs in other methods for sensitivity analysis. See \citet{zha19} for a related discussion in the context of inverse probability weighted estimators.
\section{A concordant mode of inference with additional restrictions}\label{sec:Dbar}
\subsection{A statistic with known worst-case expectation under constant effects}\label{sec:concord}
Consider instead the weighted average 
$\bar{D}^{(\tau_0)}_{\Gamma} = \sum_{i=1}^B (n_i/N){D}^{(\tau_0)}_{\Gamma i}$, where ${D}^{(\tau_0)}_{\Gamma i}$ is defined in (\ref{eq:D}). This weights each set's contribution as is typical in finely stratified experiments for inference on the sample average effect, and does not modify the value of $\Gamma$ based upon $n_i$. In light of (\ref{eq:K}) and Proposition \ref{prop:equiv}, we see that this corresponds to choosing $\kappa_{\Gamma i} = \{n_i(1+\Gamma)\}/2$ in (\ref{eq:interval}), which is equivalent to Rosenbaum's model when $n_i=2$ but is a strict subset of said model when $n_i>2$. 

\begin{theorem}\label{thm:bound}
Let $\bu$ be any vector of unmeasured confounders and suppose (\ref{eq:sensmodel}) holds at $\Gamma$. Then, if $H_F^{(\tau_0)}$ holds, $E_{\bu}(\bar{D}^{(\tau_0)}_\Gamma \mid \cF, \cZ) \leq 0$, and $E_{\bu}(\bar{D}^{(\tau_0)}_\Gamma \mid \cF, \cZ)  = 0$ for $u_{ij} = \1(\delta_{ij} \geq \tau_0)$. If instead $H_N^{(\tau_0)}$ holds but $H_F^{(\tau_0)}$ does not, we have
\begin{align}\label{eq:biasbound}
E_{\bu}(\bar{D}^{(\tau_0)}_\Gamma \mid \cF, \cZ) &\leq \sum_{i=1}^B(n_i/N)E_{\bu_i}({D}^{(\bar{\tau}_i)}_{\Gamma i}\mid \cF, \cZ)\\ 
&+ \frac{2\Gamma}{1+\Gamma}\sum_{i=1}^B(n_i/N)\left\{1 + \frac{1-\Gamma}{\Gamma}\P_{\bu_i}(\hat{\tau}_i \geq \bar{\tau}_i\mid \cF, \cZ)\right\}(\bar{\tau}_i-\tau_0)\nonumber, 
\end{align}
where $\sum_{i=1}^B(n_i/N)E_{\bu_i}({D}^{(\bar{\tau}_i)}_{\Gamma i}\mid \cF, \cZ) \leq 0$, with equality when $u_{ij} = \1(\delta_{ij} \geq \bar{\tau}_i)$.
\end{theorem}

That the worst-case expectation is bounded by zero under $H_F^{(\tau_0)}$ follows from Proposition \ref{prop:zero}, while the general form is proved in the appendix. For matched designs besides pair matching, the right-hand side of (\ref{eq:biasbound}) under the weak null need not be less than or equal to zero even if the sample average treatment effect equals $\tau_0$. The inequality (\ref{eq:biasbound}) may be strict, sometimes by a considerable margin as the simulations in \S \ref{sec:sim} reveal, such that the right-hand side being larger than zero does not imply that the worst-case expectation itself lies above zero. Nonetheless, the inequality provides a useful necessary condition for the worst-case expectation to lie above zero.



\begin{corollary}\label{cor:necessary}
Let $\bu$ be the true, but unknowable, vector of unmeasured confounders and let $\bu^*$ be the vector of unmeasured confounders yielding the worst-case expectation. Suppose (\ref{eq:sensmodel}) holds at $\Gamma$ and that $\bar{\tau} = \tau_0$. Then, a necessary condition for $E(\bar{D}_\Gamma^{(\tau_0)}\mid \cF, \cZ) > 0$ is that the following inequalities both hold:
\begin{align*}
\text{cov}\left\{\P_{\bu_i}(\hat{\tau}_i \geq \bar{\tau}_i\mid \cF, \cZ), n_i(\bar{\tau}_i - \tau_0) \right\} < 0,\;\;
\text{cov}\left\{\P_{\bu^*_i}(\hat{\tau}_i \geq \bar{\tau}_i\mid \cF, \cZ), n_i(\bar{\tau}_i - \tau_0) \right\} < 0,
\end{align*}
where $\text{cov}(\cdot)$ is the usual sample covariance.
\end{corollary}
Comparing the bounds in Theorem \ref{thm:bound} for $\bar{D}_\Gamma^{(\tau_0)}$ and Proposition \ref{prop:Tbar} for $\bar{T}_\Gamma^{(\tau_0)}$, the permutational $t$ test,  reveals a fundamental difference in the conditions required for the bias to be controlled under the weak null. When considering a sensitivity analysis using the permutational $t$-statistic as in \S\S\ref{sec:cons}-\ref{sec:challenge}, one would need to impose conditions on the relationship between individual-level treatment effects $\tau_{ij}$ and individual-level unmeasured confounders $u_{ij}$. This appears unattractive, as many natural patterns of unmeasured confounding such as essential heterogeneity directly link individual-level treatment effects to their unmeasured confounders. In contrast, the assumptions required for the bias of $\bar{D}_\Gamma^{(\tau_0)}$ to be controlled concern the relationship between the {average} of the treatment effects in a set and aggregate function of the unmeasured confounders in that set. We now provide further insight into what must occur for the bias of $\bar{D}_\Gamma^{(\tau_0)}$ not to be bounded above by zero.

\subsection{Interpreting the necessary condition under a particular pattern of adversarial bias}\label{sec:interpret}

Suppose that nature chooses the worst-case unmeasured confounder based not upon the postulated value for $\bar{\tau} = \tau_0$ being tested, but rather by maximizing the expectation of ${D}_{\Gamma i}^{(\bar{\tau}_i)}$ in each matched set $i$. Let $\bu^{**}$ denote the corresponding pattern of hidden bias. From Proposition \ref{prop:zero}, the resulting worst-case confounder has an intuitive form: $u_{ij}^{**} = \1(\delta_{ij} \geq \bar{\tau}_i) = \1\{r_{Tij} + r_{Cij}/(n_i-1) \geq \bar{r}_{Ti} + \bar{r}_{Ci}/(n_i-1)\}$. That is, in each matched set $i$, treatment assignments for which $\hat{\tau}_i \geq \bar{\tau}_i$, are given higher probability. Any choice of $\bu_i$ that differs from $\bu_{i}^{**}$ must necessarily ascribe higher probability to observing a treated-minus-control difference in means that falls {below} $\bar{\tau}_i$. This choice always maximizes $\P_{\bu_i}(\hat{\tau}_i\geq \bar{\tau}_i)$, and corresponds to the worst-case confounder for $E({D}_{\Gamma i}^{(\tau_0)}\mid \cF, \cZ)$ under $H_N^{(\tau_0)}$ for matched pairs. 

The necessary condition for $E_{\bu^{**}}(\bar{D}^{(\tau_0)}_\Gamma\mid \cF, \cZ) > 0$ under $H_N^{(\tau_0)}$ can be expressed as
\begin{align}\label{eq:nearcov}
\text{cov}\left\{\frac{n_i}{\sum_{j=1}^{n_i}\{\1(\delta_{ij}< \bar{\tau}_i)+ \Gamma\1(\delta_{ij}\geq \bar{\tau}_i)\}}, n_i(\bar{\tau}_i - \tau_0)\right\} > 0.
\end{align}The first term in the covariance (\ref{eq:nearcov}) is a function of the stratum size $n_i$ and $\sum_{i=1}^{n_i}\1(\delta_{ij} \geq \bar{\tau}_i)$, the number of potential treated-minus-control differences in means that exceed the average treatment effect in each stratum. Imagine conducting a linear regression of $\bdelta = (\delta_{11},\delta_{12},...,\delta_{Bn_B})^T$ on an $N\times B$ matrix whose $i$th column contains an indicator of membership in the $i$th matched set; call the resulting matrix $\bX$.  Let $\bH = \bX(\bX^T\bX)^{-1}\bX^T$ be corresponding hat matrix, such that $\bH\bdelta$ are the fitted values and $(\boI-\bH)\bdelta$ are the residuals. The fitted value for individual $ij$ would be $(\bH\bdelta)_{ij} = \bar{\tau}_i$, while the residual for individual $ij$ would be $\{(\boI - \bH)\bdelta\}_{ij} = \delta_{ij} - \bar{\tau}_i$. The first term in (\ref{eq:nearcov}) is a function of these residuals, while the second term is a function of the fitted values. By standard results from multiple regression, we have that $\text{cov}\{\bH\bdelta, (\boI-\bH)\bdelta\} = 0$. 
The expression (\ref{eq:nearcov}) involves a covariance between functions of $\bH\bdelta$ and functions of $(\boI-\bH)\bdelta$. This serves to highlight what must occur in order for $E_{\bu^{**}}(\bar{D}_\Gamma^{(\tau_0)}\mid \cF, \cZ) > 0$: despite the fact that the residuals are uncorrelated with the fitted values, the functions of the residuals and functions of the fitted values represented in (\ref{eq:nearcov}) must be positively associated. In particular, the proportion of positive residuals in matched set $i$ must carry information about the fitted value in matched set $i$.  This can occur when the degree of skewness in the distribution of $\delta_{ij}$ in each matched set varies as a function of $(\bar{\tau}_i-\tau_0)$. For $n_i>2$, $n_i^{-1}\sum_{i=1}^{n_i}\1(\delta_{ij}\geq \bar{\tau}_i)$ will tend to be larger under left-skewness than under right-skewness, so if $\delta_{ij}$ are right-skewed in matched sets for which $\bar{\tau}_i - \tau_0 \geq 0$ and left-skewed otherwise, the covariance in (\ref{eq:nearcov}) may be positive. While mathematically possible, it remains to be seen whether patterns of hidden bias of this form are of practical concern.

\subsection{Validity of $\bar{D}^{(\tau_0)}_\Gamma$ with heterogeneous effects for a subset of Rosenbaum's model}\label{sec:subset}
\S \ref{sec:interval} describes a general construction of test statistics that bound the worst-case expectation under $H_N^{(\tau_0)}$ under interval restrictions on the conditional probabilities. We now show that the test statistic $\bar{D}^{(\tau_0)}_\Gamma$ has expectation bounded above by zero under a particular subset of Rosenbaum's model. Consider the restriction
\begin{align}\label{eq:int2}
\frac{2}{n_i(1+\Gamma)} \leq \varrho_{ij} \leq  \frac{2\Gamma}{n_i(1+\Gamma)}.
\end{align}
When $n_i=2$,  (\ref{eq:int2}) is equivalent to the original sensitivity model (\ref{eq:sensmodel}). For $n_i>2$, this restriction is strict subset of Rosenbaum's model at $\Gamma$. Applying Propositon \ref{prop:interval} shows that $E({D}_{\Gamma i}^{(\tau_0)}\mid \cF, \cZ) \leq \bar{\tau}_i - \tau_0$ under (\ref{eq:int2}), such that $E(\bar{D}_\Gamma^{(\tau_0)}\mid \cF, \cZ)\leq 0$ when $\bar{\tau} = \tau_0$. 

The value $\kappa_{\Gamma i} = \{n_i(1+\Gamma)\}/2$ has a natural, albeit heuristic, interpretation in light of the structure of the worst-case unmeasured confounder. As highlighted in \S \ref{sec:cons}, results from \citet{ros90} show that the search for the worst-case unmeasured confounder $\bu^*$ can be reduced to an optimization over $n_i-1$ binary vectors. The $a$th of $n_i-1$ candidate vectors has $a$ ones and $n_i-a$ zeroes, resulting in $\sum_{j=1}^{n_i}\exp(\gamma u^*_{ij}) = \Gamma a + (n_i-a)$. Under the weak null we know that be one of these $n_i-1$ binary vectors will yield the worst-case expectation, but we do not know which \textit{particular} binary vector within the set attains the worst-case unless $n_i=2$. The interval (\ref{eq:intervalrosorig}) uses different values for $u_{i}^*$ when forming the upper and lower bounds on $\varrho_{ij}$, with $\sum_{j=1}^{n_i}\exp(\gamma u^*_{ij}) = \Gamma(n_i-1)+1$ in the lower bound and $\sum_{j=1}^{n_i}\exp(\gamma u^*_{ij}) = (n_i-1)+\Gamma$ in the upper bound. If we were to instead ascribe equal probability to the worst-case confounder being any of these $n_i-1$ candidate vectors, we would attain $E\{\sum_{j=1}^{n_i}\exp(\gamma u^*_{ij})\} = \{n_i(1+\Gamma)\}/2$.

\section{Variance estimation and performing the sensitivity analysis}\label{sec:inference}
\subsection{Constructing conservative standard errors for sensitivity analysis}\label{sec:SE}
Little attention has been paid until now to the variance of $\tilde{K}^{(\tau_0)}_{\Gamma}$ in \S \ref{sec:valid} or of $\bar{D}^{(\tau_0)}_{\Gamma}$ in \S \ref{sec:concord}; rather, the discussion has focused on the the extent to which the worst-case expectations of these test statistics may be understood and bounded under $H_N^{(\tau_0)}$. Here we show that the variance estimators for finely stratified designs developed in \citet{fog18mit} also yield conservative variance estimators for use in sensitivity analysis with heterogeneous effects. We cater the following construction to $\bar{D}^{(\tau_0)}_\Gamma$, but the analogous result holds for $\tilde{K}^{(\tau_0)}_\Gamma$. Let $\bQ$ be any $B\times p$ matrix that is constant over $\bz \in \Omega$ with $B>p$, and let $\bH_{\bQ}=\bQ(\bQ^T\bQ)^{-1}\bQ^T$ be its hat matrix. Let ${Y}_{\Gamma i} = B(n_i/N){D}^{(\tau_0)}_{\Gamma i}/\sqrt{1-h_{Qii}}$ where $h_{Qii}$ is the $\{i,i\}$ element of $\bH_{\bQ}$, and let $Y_\Gamma = (Y_{\Gamma 1},...,Y_{\Gamma B})^T$. Consider the standard error $\text{se}^2_\bQ(\bar{D}^{(\tau_0)}_\Gamma) = {\bY}_\Gamma^T(\boI-\bH_\bQ){\bY}_\Gamma/B^2$.
\begin{proposition}\label{prop:conservative}Regardless of the true value of $\Gamma$ for which (\ref{eq:sensmodel}) holds,
\begin{align*}
E\{\text{se}_\bQ^2(\bar{D}^{(\tau_0)}_\Gamma)\mid \cF, \cZ\} - \text{\var}(\bar{D}^{(\tau_0)}_\Gamma\mid \cF, \cZ)&= \frac{1}{B^2}E({\bY}_\Gamma\mid \cF, \cZ)^T(\boI-\bH_\bQ)E({\bY}_\Gamma\mid \cF, \cZ) \geq 0.
\end{align*}
Further, under suitable regularity conditions,
${\text{se}_\bQ^2(\bar{D}^{(\tau_0)}_\Gamma)}/\{{\text{\var}(\bar{D}^{(\tau_0)}_\Gamma\mid \cF, \cZ)}\}$ converges in probability to a value greater than or equal to one.
\end{proposition}
Proposition \ref{prop:conservative} is a straightforward extension of Proposition 1 and Theorem 2 in \citet{fog18mit}, and the proof is omitted. Despite not knowing $\bu$, Proposition \ref{prop:conservative} nonetheless provides a conservative standard error. For matched designs, a natural choice for $\bQ$ would be the vector containing weights $B(n_i/N)$ in the $i$th entry; this choice is used in the simulations in \S\ref{sec:sim}. The upwards bias of the standard errors is not reflective of a deficiency in the estimators, but rather is a fundamental feature of variance estimation under the finite population model: in general, it is impossible to consistently estimate the variance  when effects are heterogeneous \citep{din17}. 

\subsection{Large-sample reference distributions for sensitivity analysis}\label{sec:sufficient}

Suppose that $H_N^{(\tau_0)}$ is true and that (\ref{eq:sensmodel}) holds at $\Gamma > 1$. From developments in \S \ref{sec:valid}, we know that $\max_{\bu\in \mathcal{U}}\;E_{\bu}(\tilde{K}^{(\tau_0)}_\Gamma \mid \cF, \cZ) \leq 0$ without any further assumptions. From \S \ref{sec:concord} either of the following conditions are sufficient for $\max_{\bu\in \mathcal{U}}\;E_{\bu}(\bar{D}^{(\tau_0)}_\Gamma \mid \cF, \cZ) \leq 0$ in the limit:

\begin{itemize}
\item[(a)]  The true hidden bias $\bu$ satisfies $\liminf_{B\rightarrow \infty}\text{cov}\left\{\P_{\bu_i}(\hat{\tau}_i \geq \bar{\tau}_i\mid \cF, \cZ), n_i(\bar{\tau}_i - \tau_0) \right\} \geq 0$.
\item[(b)]  The hidden bias yielding the worst-case expectation $\bu^*$ satisfies\\ $\liminf_{B\rightarrow \infty}\text{cov}\left\{\P_{\bu^*_i}(\hat{\tau}_i \geq \bar{\tau}_i\mid \cF, \cZ), n_i(\bar{\tau}_i - \tau_0) \right\} \geq 0$. 
\end{itemize}Consider a candidate level-$\alpha$ sensitivity analysis using $\bar{D}^{(\tau_0)}_\Gamma$ through 
\begin{align*}
\varphi^{(\tau_0)}(\alpha,\Gamma) = \1\left\{\frac{\bar{D}^{(\tau_0)}_\Gamma}{\text{se}_\bQ(\bar{D}^{(\tau_0)}_\Gamma)}\geq \Phi^{-1}(1-\alpha)\right\},
\end{align*}
and let $\tilde{\varphi}^{(\tau_0)}(\alpha,\Gamma)$ be the analogous test based upon $\tilde{K}^{(\tau_0)}_\Gamma$. The following theorems summarize our findings for sensitivity analyses using $\bar{D}^{(\tau_0)}_\Gamma$ and $\tilde{K}^{(\tau_0)}_\Gamma$.

\begin{theorem}\label{thm:inference2}
Suppose (\ref{eq:sensmodel}) holds at $\Gamma$, and consider any $\alpha\leq 0.5$. If the weak null $H_N^{(\tau_0)}$ is true, and any of the sufficient conditions (a) or (b) hold, then under suitable regularity conditions $\limsup_{B\rightarrow \infty} E\{\varphi^{(\tau_0)}(\alpha,\Gamma)\mid \cF, \cZ\} \leq \alpha$
and equality is possible. If the constant effect model $H_F^{(\tau_0)}$ holds, then sufficient conditions (a) and (b) are both satisfied, and
$\limsup_{B\rightarrow \infty} E\{\varphi^{(\tau_0)}(\alpha,\Gamma)\mid \cF, \cZ\} \leq \alpha$ with equality possible in any matched design.
\end{theorem}

\begin{theorem}\label{thm:inference1}
Suppose (\ref{eq:sensmodel}) holds at $\Gamma$, and consider any $\alpha\leq 0.5$. If the weak null $H_N^{(\tau_0)}$ is true, then under suitable regularity conditions
${\limsup}_{B\rightarrow\infty}E\{\tilde{\varphi}^{(\tau_0)}(\alpha,\Gamma)\mid \cF, \cZ\} \leq \alpha,$ and equality is possible. If the constant effect model $H_F^{(\tau_0)}$ also holds, $\Gamma > 1$, and we do not have a paired design,
${\limsup}_{B\rightarrow\infty}E\{\tilde{\varphi}^{(\tau_0)}(\alpha,\Gamma)\mid \cF, \cZ\} < \alpha$, the inequality being strict.
\end{theorem} For paired observational studies or if $\Gamma=1$, $\tilde{K}^{(\tau_0)}_\Gamma$ and $\bar{D}^{(\tau_0)}_\Gamma$ are equivalent, implying no divergence between these modes of inference. For general matched designs with $\Gamma>1$, sensitivity analyses based upon $\tilde{K}^{(\tau_0)}_\Gamma$ are valid with heterogeneous effects, but are unduly conservative for constant effects. Sensitivity analyses with $\bar{D}^{(\tau_0)}_\Gamma$ are valid and asymptotically sharp if treatment effects are constant, but require additional, potentially plausible, assumptions to guarantee validity if treatment effects are instead heterogeneous. 

Regularity conditions on $\cF$ are discussed in the appendix. They are needed ensure that a central limit theorem holds for our test statistics, and that our estimated standard errors have limits in probability; given these, the proofs are straightforward. In the appendix, we also describe a modification of $\varphi^{(\tau_0)}(\alpha,\Gamma)$ which replaces critical values from a standard normal with critical values from a biased randomization distribution. At $\Gamma=1$, this modification results in a test that is both exact for $H_F^{(\tau_0)}$ and asymptotically correct for $H_N^{(\tau_0)}$.

\section{Bounding the worst-case expectation with binary outcomes}\label{sec:binary}
For continuous potential outcomes, it is generally impossible to explicitly calculate\\ 
$\max_{\bu\in \mathcal{U}}E_{\bu}(\bar{D}^{(\tau_0)}_\Gamma \mid \cF, \cZ)$ over $H_N^{(\tau_0)}$ despite that $\max_{\bu\in \mathcal{U}}E_{\bu}(\bar{D}^{(\tau_0)}_\Gamma \mid \cF, \cZ)=0$ under $H_F^{(\tau_0)}$ when (\ref{eq:sensmodel}) holds at $\Gamma$. With binary outcome variables the model of constant effects is particularly unsavory, as generally the onyl feasible value for a constant effect would be $\tau_0=0$. Fortunately, with binary outcomes it is possible to bound the worst-case expectation for $\bar{D}_\Gamma^{(\tau_0)}$ under $H_N^{(\tau_0)}$ when (\ref{eq:sensmodel}) is assumed to hold at $\Gamma$ using the approach of \citet{fog17}. That work develops an integer program for calculating the worst-case expectation for a wide range of test statistics suitable for binary outcomes under the sensitivity model (\ref{eq:sensmodel}). The approach proceeds by finding the worst-case expectation over all values for $\bu$ and all values for the missing potential outcomes such that $H_N^{(\tau_0)}$ holds. The formulation is such that the number of decision variables does {not} scale linearly in $N$, the number of observations. Instead, the problem scales in accord with the number of unique observed $2\times 2$ tables for each $n_i$, with treatment status on one margin and outcome on the other. For each of these observed $2\times 2$ tables, the formulation considers all possible $2\times 2$ tables with potential outcomes under treatment on one margin and under control on the other. For each table, one can then calculate the vector $\bu_i$ yielding the worst-case expectation through the separable algorithm in \S \ref{sec:cons}. 

Let $\mathtt{IP}_\Gamma^{(\tau_0)}(\bR, \bZ)$ be the solution to this integer program, which upper bounds \\$\max_{\bu\in \mathcal{U}}E_{\bu}(\bar{D}^{(\tau_0)}_\Gamma \mid \cF, \cZ)$. Consider the candidate sensitivity analysis\\
$\varphi_{binary}^{(\tau_0)}(\alpha,\Gamma) = \1\left[\{\bar{D}^{(\tau_0)}_\Gamma - \mathtt{IP}_\Gamma^{(\tau_0)}(\bR, \bZ)\}/\{\text{se}_\bQ(\bar{D}^{(\tau_0)}_\Gamma)\}\geq \Phi^{-1}(1-\alpha)\right]$.
\begin{theorem}\label{thm:binary}
Suppose (\ref{eq:sensmodel}) holds at $\Gamma$ and that outcomes are binary. If $H_N^{(\tau_0)}$ holds, then under suitable regularity conditions 
${\limsup}_{B\rightarrow\infty} E\{\varphi_{binary}^{(\tau_0)}(\alpha,\Gamma)\mid \cF, \cZ\} \leq \alpha.$\end{theorem}
Consider a paired observational study. There are four potentially observed $2\times 2$ tables corresponding to the possible combinations of $(Z_{ij}, R_{ij})$ that could be observed in each pair. For each of these observed tables, there are four possible $2\times 2$ tables of potential outcomes based on the potential values for the unobserved potential outcomes. This results in a total of at most 16 decision variables for {any} sample size $N$. Had we instead encoded the unobserved binary potential outcome as the decision variables there would have been $N$ decision variables, rendering the problem infeasible for moderately sized $N$.  See \citet[][ \S\S 4-5]{fog17} for a detailed description of the integer program, and for computational experiments highlighting the strength of the underlying formulation.

\section{Simulations with worst-case hidden bias}\label{sec:sim}
\subsection{The generative model}
We compare the candidate modes of sensitivity analyses with continuous potential outcomes generated from varied distributional shapes. Similar simulations comparing our methods for binary outcomes are contained in the appendix. There are $B = 500$ matched sets in the $m$th of $M=5000$ iterations, and matched set sizes are drawn independently and identically distributed with $n_i\sim 2 + Poisson(2)$ to mimic an observational study using variable ratio matching. In each matched set, exactly one individual receives the treatment and the remaining $n_i-1$ receive the control. In the $m$th iteration, potential outcomes are drawn independently for distinct individuals as
\begin{align}\label{eq:gen2}
{r}_{Cij}  = \varepsilon_{Cij};\;\;
{r}_{Tij} = {r}_{Cij} + \beta_{i} + \varepsilon_{Tij},
\end{align} 
where $\beta_i$ are independent and identically distributed according to some $F_{\beta}(\cdot)$, $\varepsilon_{Cij}$ are independently distributed according to $F_{Ci}(\cdot;\beta_i)$, $\varepsilon_{Tij} $ are independent according to $F_{T}(\cdot;\beta_i)$, $\varepsilon_{Cij}$ and $\varepsilon_{Tij}$ are independent, and $E(\beta_i) = E(\varepsilon_{Tij}) = E(\varepsilon_{Cij}) = 0$. The variance of $\beta_i$ affects the across-set effect heterogeneity (potentially reflecting the impact of effect modifiers $\bx_i$), while $\varepsilon_{Tij}$ affects the within-set effect heterogeneity. Note the potential dependence of $F_{C}(\cdot;\beta_i)$ and $F_{T}(\cdot;\beta_i)$ on $\beta_i$. We then construct the three vectors of unmeasured confounders resulting in the worst-case expectations for $\tilde{K}^{(\tau_0)}_\Gamma$, $\bar{D}^{(\tau_0)}_\Gamma$ and $\bar{T}^{(\tau_0)}_\Gamma$  when (\ref{eq:sensmodel}) holds at $\Gamma=5$ and with $\tau_0 = \bar{\tau}^{(m)}$. For each test, we generate a single treatment assignment vector through (\ref{eq:sensmodel}) using the worst-case vector of unmeasured confounders for that test. We finally conduct the sensitivity analysis at $\Gamma$ for the null hypothesis $\bar{\tau} = \bar{\tau}^{(m)}$.


\begin{table}
\caption{\label{tab:gen}Generative models for the simulation study. $\mathcal{N}(\mu, \sigma^2)$ is a normal with mean $\mu$ and variance $\sigma^2$, $\tilde{\mathcal{E}}(\lambda) = \mathcal{E}(\lambda) - 1/\lambda$ is an exponential with rate $\lambda$ minus its expectation, and $V_i  = \{2\1(\beta_i\geq 0)-1\}$. }
\centering
\begin{tabular}{llll llll}
          	&	       $\beta_i$       	&	       $\varepsilon_{Cij}$     	&	       \multicolumn{1}{l}{$\varepsilon_{Tij}$}        	&	               	&	       \multicolumn{1}{l}{$\beta_i$}       	&	       $\varepsilon_{Cij}$     	&	       $\varepsilon_{Tij}$    \\
(a)      	&	0	&	       $\tilde{\mathcal{E}}(1/10)$     	&	0	&	(f)  	&	 $-\tilde{\mathcal{E}}(1)$ 	&	 $-\tilde{\mathcal{E}}(1/10)$	&	 $-\tilde{\mathcal{E}}(1/10)$\\ 
(b)      	&	       $\mathcal{N}(0,1)$      	&	       $\mathcal{N}(0,10^2)$   	&	        $\mathcal{N}(0,10^2)$          	&	      (g)      	&	       $-\tilde{\mathcal{E}}(1/10)$    	&	       $-\tilde{\mathcal{E}}(1/10)$    	&	       $-\tilde{\mathcal{E}}(1/10)$\\
(c)      	&	       $\mathcal{N}(0,10^2)$   	&	       $\mathcal{N}(0,10^2)$   	&	        $\mathcal{N}(0,10^2)$          	&	      (h)      	&	       $\mathcal{N}(0,1)$      	&	        $V_i\{\tilde{\mathcal{E}}(1/10)\}$     	&	       $\mathcal{N}(0,1)$\\   
(d)      	&	       $\tilde{\mathcal{E}}(1)$        	&	       $\tilde{\mathcal{E}}(1/10)$     	&	       $\tilde{\mathcal{E}}(1/10)$     	&	      (i)      	&	       $\mathcal{N}(0,5^2)$    	&	        $V_i\{\tilde{\mathcal{E}}(1/10)\}$     	&	       $\mathcal{N}(0,1)$\\   
(e)      	&	       $\tilde{\mathcal{E}}(1/10)$     	&	       $\tilde{\mathcal{E}}(1/10)$     	&	       $\tilde{\mathcal{E}}(1/10)$     	&	      (j)      	&	       $\mathcal{N}(0,1)$      	&	       $-V_i\{\tilde{\mathcal{E}}(1/10)\}$     	&	       $\mathcal{N}(0,1)$\\   
	&		&		&		&	      (k)      	&	       $\mathcal{N}(0,5^2)$    	&	        $-V_i\{\tilde{\mathcal{E}}(1/10)\}$            	&	       $\mathcal{N}(0,1)$\\
\end{tabular}
\end{table}


Table \ref{tab:gen} provides the 11 combinations of distributions for $\beta_i, \varepsilon_{Cij}$, and $\varepsilon_{Tij}$ used in the simulation. Under choice (a), $H_F^{(0)}$ holds in all simulations. Choices (b) and (c) lead to a symmetric distribution for $\delta_{ij}$ within each matched set, with more across-set treatment effect heterogeneity in (c) than (b). Choices (d) and (e) lead to right-skewed distributions for $\delta_{ij}$ with more heterogeneity of effects in (e), and (f) and (g) lead to left-skewed distributions for $\delta_{ij}$ with more heterogeneity in (g). Choices (h) and (i) result in distributions for $\delta_{ij}$ that are right-skewed when $\beta_i \geq 0$ and left-skewed otherwise, while for (j) and (k) it is reversed. Choice (i) has more across-set effect heterogeneity than (h), and (k) has more across-set heterogeneity than (j). 
\subsection{Results}


Table \ref{tab:results} contains the results of the simulation study. For each setting, we report the estimated Type I error rate. We further include the expected value of the difference between the test statistic employed and its candidate worst-case expectation, scaled by the test statistic's standard deviation across simulations. Finally, we report upper bounds given in Theorem \ref{thm:weak} for $\tilde{K}^{(\tau_0)}_\Gamma$, Theorem \ref{thm:bound} for $\bar{D}^{(\tau_0)}_\Gamma$, and Proposition \ref{prop:Tbar} for  $\bar{T}^{(\tau_0)}_\Gamma$, scaled again by the test statistic's standard deviation. The first set of columns show that $\tilde{K}^{(\tau_0)}_\Gamma$ was valid in all settings as predicted by Theorem \ref{thm:inference1}. This conservativeness was present even in setting (a) where the sharp null holds, reflecting the consequences of Theorem \ref{thm:impossible}.  The permutational $t$-test, displayed in the third set of columns, correctly controlled the size of the procedure when the sharp null held as expected, but was anti-conservative by a sizeable margin in many plausible simulation settings with heterogeneous effects. 

The second set of columns summarize findings for $\bar{D}^{(\tau_0)}_\Gamma$. As Theorem \ref{thm:inference2} predicts, under the sharp null the size was controlled at 0.10. With continuous outcomes, we see that the procedure was conservative for all settings except (h), and was appreciably less conservative than $\tilde{K}^{(\tau_0)}_\Gamma$ as predicted by Theorem \ref{thm:bound}. That the method failed for simulation (h) further highlights Theorem \ref{thm:impossible}: because the procedure is sharp under constant effects, there must be situations where the method is invalid under the weak null. The bound on the worst-case bias can be quite conservative in many simulation settings with heterogeneous effects. For settings (a)-(g) and (j)-(k) the necessary conditions in Corollary \ref{cor:necessary} for the procedure to be invalid were not satisfied, while in simulation (i) we see the procedure remained valid despite the necessary conditions being satisfied. That is, the necessary conditions may be far from sufficient.

A pattern observed throughout the simulation study is that all candidate modes of sensitivity analysis rejected in the null {less} frequently in the presence of strong across-set effect heterogeneity (captured by the variance of $\beta_i$) than in the presence of mild across-set heterogeneity.  To explain this for $\bar{D}_\Gamma^{(\tau_0)}$, the function $E_{\bu_i}({D}_{\Gamma i}^{(\tau_0)}\mid \cF, \cZ)$ is concave in $\tau_0$ for fixed values $\bu_i$, and the bound in Theorem \ref{thm:bound} stems from a Taylor expansion about the point $\bar{\tau}_i$. When effects are severely heterogeneous across matched sets but $\bar{\tau}=\tau_0$, the value $\tau_0$ may be quite far from $\bar{\tau}_i$, which in turn increases the conservativeness of the bound based upon the Taylor expansion. Extreme effect modification promotes conservativeness, and  ${D}_{\Gamma i}^{(\tau_0)}$ is least conservative when across-set effect modification is limited. 

\begin{table}
\caption{\label{tab:results}Results from the simulation. The sensitivity model (\ref{eq:sensmodel}) holds at $\Gamma=5$, and the desired size was $\alpha=0.10$. Presented are the estimated Type I error rate, along with the actual bias in the test statistic and the upper bound on the bias in units of the statistic's standard deviation.}
\centering\begin{tabular}{lccccccccc}
&\multicolumn{3}{c}{$\tilde{K}_\Gamma$} & \multicolumn{3}{c}{$\bar{D}_\Gamma$}&\multicolumn{3}{c}{$\bar{T}_\Gamma$}\\
 &Size & Bias & Bound & Size & Bias & Bound& Size &  Bias&Bound\\
(a)	&	0.000	&	-4.022	&	0.000	&	0.101	&	0.000	&	0.000	&	0.096	&	0.000	&	0.000	\\
(b)	&	0.000	&	-4.093	&	0.000	&	0.034	&	-0.649	&	-0.069	&	0.372	&	0.898	&	2.694	\\
(c)	&	0.000	&	-6.062	&	0.000	&	0.000	&	-2.764	&	-0.199	&	0.034	&	-0.389	&	2.626	\\
(d)	&	0.000	&	-4.720	&	0.000	&	0.024	&	-0.695	&	-0.085	&	0.329	&	0.848	&	2.343	\\
(e)	&	0.000	&	-6.456	&	0.000	&	0.000	&	-2.661	&	-0.115	&	0.067	&	-0.127	&	2.293	\\
(f)	&	0.000	&	-3.321	&	0.000	&	0.041	&	-0.602	&	-0.082	&	0.626	&	1.458	&	3.175	\\
(g)	&	0.000	&	-6.057	&	0.000	&	0.000	&	-3.315	&	-0.228	&	0.038	&	-0.428	&	3.164	\\
(h)	&	0.000	&	-3.157	&	0.000	&	0.138	&	0.060	&	0.113	&	0.086	&	-0.047	&	0.030	\\
(i)	&	0.000	&	-4.190	&	0.000	&	0.016	&	-0.844	&	0.512	&	0.011	&	-0.865	&	-0.049	\\
(j)	&	0.000	&	-3.508	&	0.000	&	0.091	&	-0.181	&	-0.130	&	0.098	&	0.016	&	0.086	\\
(k)	&	0.000	&	-4.980	&	0.000	&	0.001	&	-1.607	&	-0.669	&	0.037	&	-0.476	&	0.226	\\
\end{tabular}
\end{table}

\section{Data Example: Smoking and Lead}\label{sec:example}
Using data from the 2007-2008 National Health and Nutrition Examination Survey (NHANES), \citet{ros13} constructed two matched data sets comparing lead levels ($\mu$ g/dl) in the blood for daily smokers and nonsmokers. The first matched comparison contains 250 smoker-nonsmoker pairs, while the second comparison consists of 150 matched sets containing one smoker and five matched nonsmoker. In both cases, individuals were matched on gender, age, race, education level, and household income level; see the appendix of \citet{ros13} for additional details. 

Table \ref{tab:data} contains the results of the sensitivity analyses performed at $\alpha = 0.05$ on both data sets using the unstudentized permutational $t$-test along with the tests $\tilde{\varphi}^{(0)}$ and $\varphi^{(0)}$ for various values of $\Gamma$. All tests are conducted using a normal approximation. As can be seen, for the paired data set $\tilde{\varphi}^{(0)}(0.05, \Gamma)$ and $\varphi^{(0)}$ yield identical conclusions for all values of $\Gamma$. This is because when $n_i=2$, $\tilde{\kappa}_{n_i} = \Gamma + 1$ and $\Gamma_{n_i} = \Gamma$ in (\ref{eq:intervalros}), such that $\tilde{K}_\Gamma^{(0)} =\bar{D}_\Gamma^{(0)}$. By Theorem \ref{thm:inference2} in concert with having a paired design, $\tilde{\varphi}^{(0)}= \varphi^{(0)}$ yields inference that is generally asymptotically conservative, but limiting size equal to $\alpha$ is possible; see also Theorem 1 of \citet{fog19student}. In contrast, the large-sample approximation for the permutational $t$-test yields a valid sensitivity analysis for Fisher's sharp null at any $\Gamma$ and is valid for the weak null at $\Gamma=1$, but may be anti-conservative even asymptotically for the weak null; see Theorem 4 of \citet{fog19student}. The largest values of $\Gamma$ for which the tests rejected the null hypothesis were 1.87 for the permutational $t$, and 2.01 for $\tilde{\varphi}^{(0)}= \varphi^{(0)}$. Not only does the permutational $t$ generally run the risk of being anti-conservative for some allocations of potential outcomes, but here the reported insensitivity to hidden bias was \textit{lower} for the permutational $t$ than it was for the asymptotically valid test $\tilde{\varphi}^{(0)}= \varphi^{(0)}$.

\begin{table}
\begin{center}
\caption{\label{tab:data} Worst-case $p$-values for smoking and lead. For each matched data set, sensitivity analyses are conducted using the permutational $t$-test along with $\tilde{\varphi}^{(0)}$ and $\varphi^{(0)}$ for different values of $\Gamma$. For each method, candidate worst-case $p$-values are reported.}
\begin{tabular}{r c c c c c c c}
&\multicolumn{3}{c}{1:1 matching, $B=250$} & \multicolumn{3}{c}{1:5 matching, $B=150$}\\
$\Gamma$ & Perm. $t$ & $\tilde{\varphi}^{(0)}$ & $\varphi^{(0)}$ & Perm. $t$. & $\tilde{\varphi}^{(0)}$ & $\varphi^{(0)}$ \\
1.00& 0.0000& 0.0000&0.0000& 0.0009 & 0.0026 & 0.0026 \\
1.25 & 0.0007& 0.0002&0.0002&0.0114 & 0.0370 & 0.0134\\
1.50 & 0.0067& 0.0023&0.0023&0.0519 & 0.1998& 0.0460\\
1.75 & 0.0291& 0.0131&0.0131&0.1362 & 0.4919 & 0.1136\\
2.00 & 0.0807& 0.0478&0.0478&0.2574 & 0.7467 & 0.2178\\
\end{tabular}
\end{center}
\end{table}

For the 1:5 match, $\tilde{\varphi}^{(0)}$ and $\varphi^{(0)}$ are no longer equivalent beyond $\Gamma=1$: $\tilde{\varphi}^{(0)}$ uses a larger worst-case expectation to guarantee asymptotic conservativeness over the entirety of the weak null, whereas $\varphi^{(0)}$ provides conservative inference over the subset of the weak null where either conditions (a) or (b) of \S \ref{sec:sufficient} holds. We see in Table \ref{tab:data} that this can have a large impact on the performance of the sensitivity analysis, as the worst-case $p$-values for $\tilde{\varphi}^{(0)}$ are markedly larger than those of $\varphi^{(0)}$. This is also reflected in the corresponding changepoint $\Gamma$ values for inference at $\alpha=0.05$, which are 1.29 and 1.52 for $\tilde{\varphi}^{(0)}$ and $\varphi^{(0)}$ respectively. For this data set, the permutational $t$ performs similarly to the procedure $\varphi^{(0)}$, with a changepoint $\Gamma$ of 1.49. 

In the 1:5 match, inference for the weak null using the unstudentized permutational $t$ cannot generally be trusted even at $\Gamma=1$ due to the treated-control imbalance. The simulation studies in \S \ref{sec:sim} indicate that sensitivity analyses using permutational $t$ can fail disastrously under many realistic data generating models where only the weak null holds; see the Supplementary Materials for a further illustration. This leads us to discourage the use of the permutational $t$ for sensitivity analysis for the weak null in general matched designs. While $\tilde{\varphi}^{(0)}$ provides asymptotically valid inference over the entirety of the weak null, it does so at a steep price in terms of reported insensitivity to hidden bias. The procedure $\varphi^{(0)}$ is not valid over all elements of the weak null, but the discussion in \S \ref{sec:Dbar} and simulations in \S\ref{sec:sim} suggest that the elements of the weak null and patterns of hidden bias giving rise to anti-conservative sensitivity analyses through $\varphi^{(0)}$ may be of limited practical concern. Only setting (h) in the simulation study produced anti-conservative inference, whereas those settings reflecting more conventional generative models failed to break $\varphi^{(0)}$.

\section{Discussion} 
For pair matching, the choice of sensitivity analysis for testing the weak null is straightforward: $\tilde{K}^{(\tau_0)}_\Gamma$, $\bar{D}^{(\tau_0)}_\Gamma$ and $\bar{T}^{(\tau_0)}_\Gamma$ are all equivalent, and the studentized sensitivity analysis  described in detail in \citet{fog19student} using these statistics provides a sensitivity analysis for the weak null that remains sharp if effects are constant. For testing weak nulls in more flexible matched designs, the researcher must instead weigh the benefits and downsides of a few competing methods. If bounding the worst-case expectation over the entirety of the weak null and for all patterns of hidden bias is deemed essential and non-negotiable, the sensitivity analysis using $\tilde{K}_\Gamma^{(\tau_0)}$ is appropriate. While there exist numerical examples yielding exactness rather than conservativeness for this test, they may be viewed as pathological. The sensitivity analysis can be very conservative under reasonable data-generating processes such as those used in \S \ref{sec:sim}, and can yield markedly lower reported insensitivites to hidden biases as the data example in \S \ref{sec:example} highlighted. 

The statistic $\bar{D}_\Gamma^{(\tau_0)}$ employs an upper bound on the worst-case expectation that is tight under constant effects. The necessary conditions within Corollary \ref{cor:necessary}, in concert with the discussion in \S\ref{sec:interpret}, give a sense of what must go wrong for $\bar{D}_\Gamma^{(\tau_0)}$ to fail: despite residuals being uncorrelated with fitted values in linear regression, a function of the residuals must be strongly correlated with a function of the fitted values. Those considerations were exploited in simulation setting (h), but it may be difficult to imagine a pattern of hidden bias affecting an observational study in this way being of scientific interest. The procedure $\varphi^{(\tau_0)}(\alpha,\Gamma)$ provides a unified mode of inference for Fisher's null hypotheses and the subset of Neyman's null satisfying either of the sufficient conditions in \S \ref{sec:sufficient} for observational studies using any optimal without-replacement matching algorithm. 
\appendix
\newpage
\begin{center}\LARGE{Appendix}\end{center}

\section{Illustrating the pitfalls of the permutational $t$}
\subsection{The setup}
We illustrate through simulated data the answers the following questions:
\begin{description}\item[(Q1)] Assuming $\Gamma=1$ as would be the case in a finely stratified experiment, does inference assuming effects are constant at $\tau_0$ and using $\hat{\tau}$ as a test statistic control the Type I error rate in the limit if instead effects are heterogeneous?
\item[(Q2)] When conducting a sensitivity analysis at $\Gamma > 1$, does the worst-case expectation derived in \S \ref{sec:cons} assuming constant effects also bound the worst-case expectation when effects are heterogeneous and average to $\tau_0$?
\end{description} 

There are $B=500$ matched sets, each containing $n_i=5$ individuals, and the sensitivity model (\ref{eq:sensmodel}) holds at some value $\Gamma = \exp(\gamma)\geq 1$. In each matched set, exactly one individual receives the treatment and the remaining $4$ receive the control. In the $m$th of $M$ iterations, potential outcomes in each matched set are drawn as
\begin{align}\label{eq:gen}
r_{Cij} \sim  \mathcal{E}(1/15);\;\;\;r_{Tij}\mid r_{Cij}  \sim r_{Cij} + \mathcal{E}(1/30),
\end{align} 
where $\mathcal{E}(\lambda)$ is an exponential distribution with rate $\lambda$. The treatment effects $\tau_{ij} = r_{Tij} - r_{Cij}$ are heterogeneous and follow an exponential distribution with rate 1/30, and the sample variances for the potential outcomes under treatment, under control, and for the treatment effects have expectations $E(\hat{\sigma}^2_{Ci}\mid \cZ) = 15^2$; $E(\hat{\sigma}^2_{Ti}\mid \cZ) = 15^2+30^2$; and $E(\hat{\sigma}^2_{\tau i}\mid \cZ) = 30^2$ respectively.  Denote by $\bar{\tau}^{(m)}$ the average of the treatment effects in iteration $m$. With the potential outcomes specified, worst-case unmeasured covariates $\bu$ are then constructed to maximize the expectation of $\hat{\tau}$ when the sensitivity model (\ref{eq:sensmodel}) holds at some value $\Gamma$, and we calculate the conditional assignment probabilities $\varrho$. Using these probabilities, we assign exactly one individual in each matched set to the treatment and four to the control, resulting in a vector of observed response $\bR^{(m)}$. Finally, we conduct a sensitivity analysis at $\Gamma$ with $\hat{\tau} - \bar{\tau}^{(m)}$ as the test statistic and assuming constant effects despite effects actually being heterogeneous. Using the separable algorithm described in \S \ref{sec:cons}, we calculate a candidate worst-case $p$-value $\tilde{p}_\Gamma\{\bar{\tau}^{(m)}\}$, asymptotically valid under the assumption of constant effects, and record whether or not we rejected the null.

\subsection{Improper variance in a finely stratified experiment}\label{sec:badvar2}
We first consider a finely stratified experiment under the generative model (\ref{eq:gen}), such that $\Gamma=1$ and $\P(Z_{ij}=1\mid \cF, \cZ) = 1/5$ in each matched set regardless of $\bu$. We address (Q1) through a comparison of the true variance of $\hat{\tau}$ across randomizations to the variance under the assumption of constant effects employed when calculating $\tilde{p}_1(\tau_0)$. In matched designs, the true variance for $\sqrt{B}\hat{\tau}$ converges in probability to\begin{align*}\underset{B\rightarrow\infty}{\text{plim}}\;\;B\sum_{i=1}^B\left(\frac{n_i}{N}\right)^2\left\{\hat{\sigma}^2_{Ci}/(n_i-1) + \hat{\sigma}^2_{Ti}- \hat{\sigma}^2_{\tau i}/n_i\right\}, 
\end{align*} 
which equals 1001.25 in the generative model (\ref{eq:gen}). Meanwhile, the reference distribution assuming constant effects uses a pooled variance of the treated and control individuals constructed under the incorrect assumption that the variances are equal. The variance for $\sqrt{B}\hat{\tau}$ used by the permutational $t$-test instead limits to
\begin{align*}
\underset{B\rightarrow\infty}{\text{plim}}\;\; B \sum_{i=1}^B\left(\frac{n_i}{N}\right)^2\left\{\frac{(n_i-1)\hat{\sigma}^2_{Ci} + \hat{\sigma}^2_{Ti}}{n_i-1}\right\},
\end{align*} which equals 506.25 in the generative model (\ref{eq:gen}). Recall that $\tilde{p}_1(\tau_0)$ is the large-sample $p$-value generated using the distribution of $\hat{\tau} - \tau_0$ under the assumption of constant effects. We then see that the limiting probability of a Type I error if we reject when $\tilde{p}_1\{\bar{\tau}^{(m)}\} \leq 0.05$ is actually $\Phi(\Phi^{-1}(0.05)\sqrt{506.25/1001.25}) = 0.12$. The answer to $(Q1)$ is negative: even at $\Gamma=1$, inference assuming constant effects and using $\hat{\tau}-\tau_0$ as the test statistic need not control the asymptotic Type I error rate.

\subsection{Improper worst-case expectation in matched observational studies} \label{sec:wrong2}
We again consider the generative model (\ref{eq:gen}), but now when (\ref{eq:sensmodel}) holds at $\Gamma=6$. Recall that the true values for the unmeasured confounders $\bu$ in our simulation equal the unmeasured confounders yielding the worst-case (largest) expectation for $\hat{\tau}-\tau_0$. Once these confounders have been set, treatments are assigned in each set based upon the resulting probabilities ${\varrho}_i$, leading to a vector of observed responses $\bR^{(m)}$. We then proceed with a sensitivity analysis at $\Gamma=6$ under the incorrect assumption of constant effects, using the separable algorithm in \S\ref{sec:cons}. This results in candidate worst-case expectations $\mu_{\Gamma i}^{(m)}$ and variances $\nu_{\Gamma i}^{(m)}$ for each $i$.  Because the constant effects model is actually false, the candidate worst-case expectation generated by the separable algorithm need not equal the true worst-case expectation.  

\begin{center}
\begin{figure}[h]
\includegraphics[scale = .5]{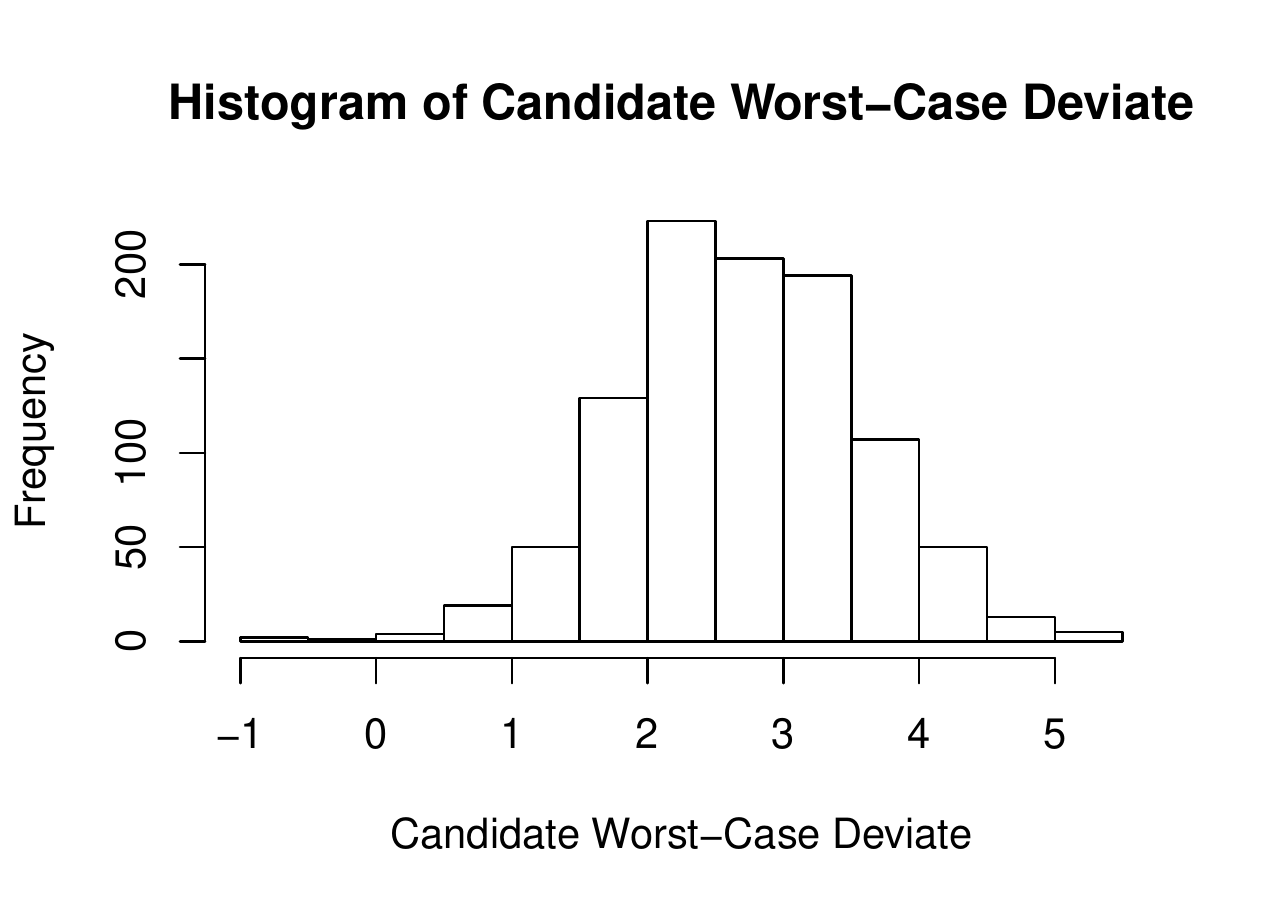}
\includegraphics[scale = .4]{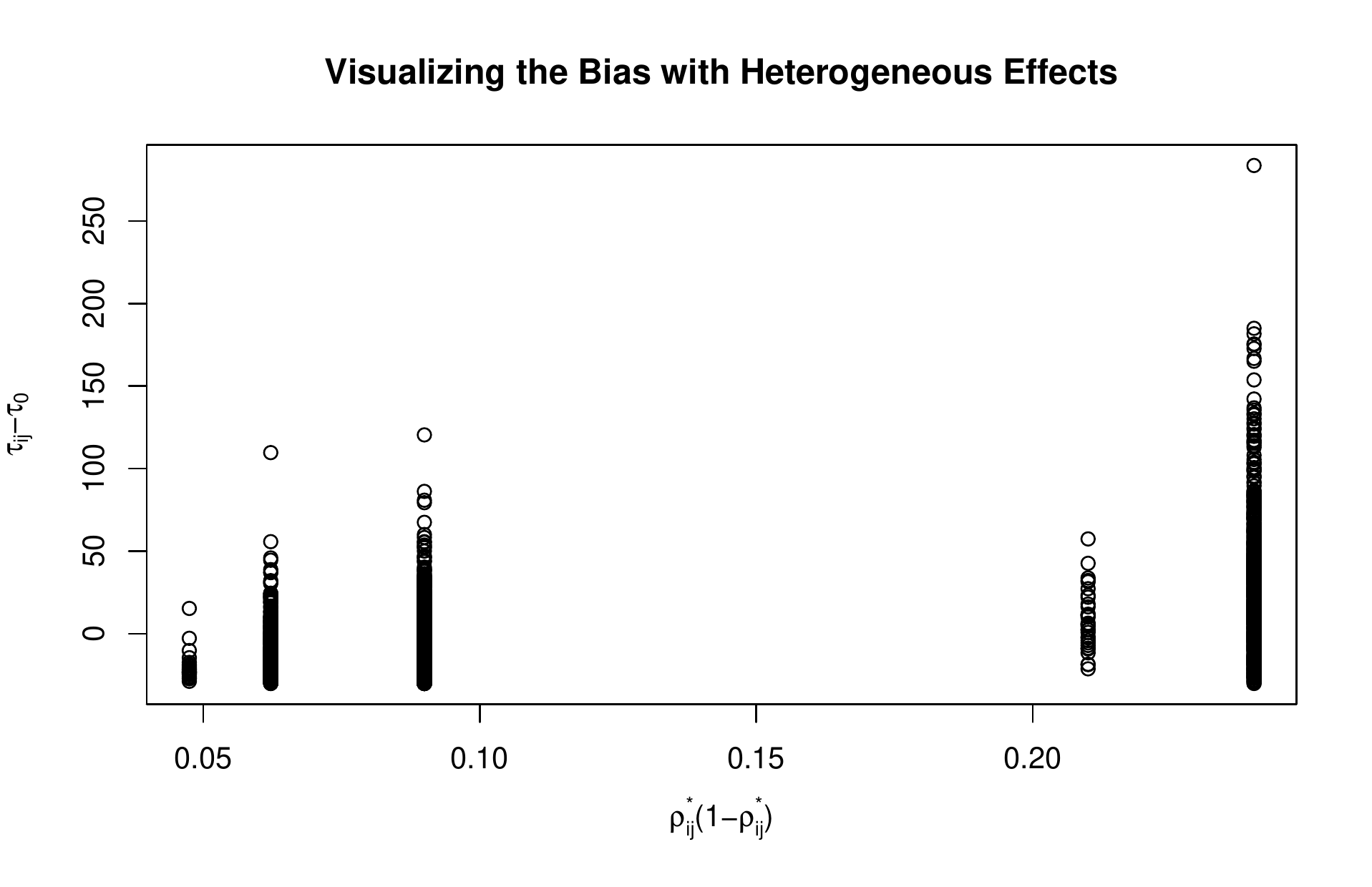}
\caption{(Left) The distribution of the candidate worst-case deviate with heterogeneous effects at $\Gamma=6$. (Right) The underlying relationship between $\varrho^*_{ij}(1-\varrho_{ij}^*)$ and $\tau_{ij} - \tau_0$.}\label{fig:asymsepfail}
\end{figure}
\end{center}
Under a Gaussian limit, it must be that $E(\hat{\tau} - \tau_0 - \sum_{i=1}^B\mu_{\Gamma i}\mid \cF, \cZ)\leq 0$ for the sensitivity analysis assuming constant effects to also be valid under effect heterogeneity should a suitable studentization be employed. Figure \ref{fig:asymsepfail} shows the distribution of the worst-case standardized deviated generated by asymptotic separability across $M=1000$ simulations, which would be asymptotically standard normal under constant effects. While it is still roughly normal with heterogeneous effects, the mean is 2.67 and the standard deviation is 0.89. That the true expectation is substantially above zero invalidates the permutational $t$-based sensitivity analysis under effect heterogeneity, which proceeds as though this distribution is standard normal. When rejecting when $\tilde{p}_\Gamma\{\bar{\tau}^{(m)}\} \leq 0.05$ in our simulation the estimated Type I error rate is 0.80, revealing that the answer to (Q2) is also negative. Figure \ref{fig:asymsepfail} shows the relationship between $\varrho_{ij}(1-\varrho_{ij})$ = $\varrho_{ij}^*(1-\varrho_{ij}^*)$ and $\tau_{ij} - \tau_0$ in a given realization of the generative model. We see that larger values for $\varrho_{ij}^*(1-\varrho_{ij}^*)$ correspond to larger values for $\tau_{ij} - \tau_0$. Proposition \ref{prop:Tbar} warns that this type of relationship may lead to a breakdown in the permutational $t$, and this possibility is realized here.

\section{Type I error simulations with binary outcomes}
\subsection{The generative model}
For binary outcomes, we compare $\tilde{\varphi}^{(\tau_0)}(\alpha,\Gamma)$, $\hat{\varphi}^{(\tau_0)}(\alpha,\Gamma)$ and $\hat{\varphi}_{binary}^{(\tau_0)}(\alpha,\Gamma)$. There are $B$ matched sets in each iteration, and matched set sizes are drawn $iid$ with $n_i\sim 2 + Poisson(2)$ to mimic an observational study using variable ratio matching. In each matched set, exactly one individual receives the treatment and the remaining $n_i-1$ receive the control. As in \S \ref{sec:sim}, in the $m$th of $M$ iterations potential outcomes are drawn independently for distinct individuals as
\begin{align}\label{eq:gen2}
\tilde{r}_{Cij} = \varepsilon_{Cij};\;\;
\tilde{r}_{Tij} = \tilde{r}_{Cij} + \beta_{i} + \varepsilon_{Tij},
\end{align} 
where $\beta_i$ are independent and identically distributed according to some $F_{\beta}(\cdot)$, $\varepsilon_{Cij}$ are independently distributed according to $F_{Ci}(\cdot;\beta_i)$, $\varepsilon_{Tij} $ are independent according to $F_{T}(\cdot;\beta_i)$, $\varepsilon_{Cij}$ and $\varepsilon_{Tij}$ are independent, and $E(\beta_i) = E(\varepsilon_{Tij}) = E(\varepsilon_{Cij}) = 0$. The variance of $\beta_i$ affects the across-set effect heterogeneity (potentially reflecting the impact of effect modifiers $\bx_i$), while $\varepsilon_{Tij}$ affects the within-set effect heterogeneity. Note the potential dependence of $F_{C}(\cdot;\beta_i)$ and $F_{T}(\cdot;\beta_i)$ on $\beta_i$. We then set $r_{Tij} = \1(\tilde{r}_{Tij} > 0)$, and set $r_{Cij} = \1(\tilde{r}_{Cij} > 0)$. The sample average treatment effect is then $N^{-1}\sum_{i=1}^B\sum_{j=1}^{n_i}(r_{Tij}-r_{Cij})=\bar{\tau}^{(m)}$.

Once the potential outcomes are fixed, we construct the two vectors of unmeasured confounders  resulting in the worst-case expectations for $\tilde{K}^{(\tau_0)}_\Gamma$ and $\bar{D}^{(\tau_0)}_\Gamma$  when (\ref{eq:sensmodel}) holds at $\Gamma$ and with $\tau_0 = \bar{\tau}^{(m)}$; these worst-case vectors of hidden covariates may be different. For each test, we generate a single treatment assignment vector through (\ref{eq:sensmodel}) using the worst-case vector of unmeasured confounders for that test. We finally conduct the sensitivity analysis at $\Gamma$ for the null hypothesis $\bar{\tau} = \bar{\tau}^{(m)}$. We use the 11 simulations settings provided in Table \ref{tab:gen} of the manuscript, and report similar performance metrics for the test statistics when deployed with binary outcomes. For the integer-programming approach in \S \ref{sec:binary} however, we instead report the average solution time as no analytical bound is available.

\subsection{Results}


Table \ref{tab:resultsbinary} contains the results of the simulation study with binary outcomes, conducting a sensitivity analysis at $\Gamma=5$ while attempting to control the Type I error rate at $\alpha = 0.1$. The results in the first two columns are similar to those observed in the continous case: the procedure in the first column is conservative in all simulation settings, while the procedure in the second set of columns is exact under Fisher's sharp null, but anti-conservative in settings (h) and (j). For binary outcomes,  the integer-programming approach correctly controlled the Type I error rate in all simulation settings, thus remedying the anti-conservativeness of $\hat{\varphi}_\Gamma^{(\tau_0)}(0.1, 5)$. Comparing the relative biases, it was less conservative than $\tilde{\varphi}^{(\tau_0)}(0.1, 5)$ in settings (a)-(f), but more conservative in (h)-(k) where skewness varied with the average treatment effects. For all simulation settings, the integer program solved in well under half a second on a personal laptop with a 3.5 GHz processor and 16.0 GB RAM using the \texttt{gurobi} package within \texttt{R}. 

\begin{table}
\caption{\label{tab:resultsbinary}Results from the simulation study with binary outcomes. The sensitivity model (\ref{eq:sensmodel}) holds at $\Gamma=5$, and the desired size was $\alpha=0.10$. Presented are the estimated Type I error rate for each method, along with the actual bias in the test statistic and the upper bound on the bias in units of the statistic's standard deviation. For the method using an integer program, solution time in seconds is instead reported.}
\centering
\begin{tabular}{lccccccccc}
&\multicolumn{3}{c}{$\tilde{K}_\Gamma$} & \multicolumn{3}{c}{$\bar{D}_\Gamma$}&\multicolumn{3}{c}{$\mathtt{IP}_\Gamma$}\\
 &Size & Bias & Bound & Size & Bias & Bound& Size &  Bias & Time (s)\\
(a)	&	0.000	&	-4.368	&	0.000	&	0.102	&	0.000	&	0.000	&	0.017	&	-0.912	&	0.283	\\
(b)	&	0.000	&	-4.850	&	0.000	&	0.005	&	-1.189	&	-0.674	&	0.001	&	-1.961	&	0.300	\\
(c)	&	0.000	&	-7.117	&	0.000	&	0.000	&	-3.412	&	-0.969	&	0.000	&	-4.182	&	0.306	\\
(d)	&	0.000	&	-5.630	&	0.000	&	0.003	&	-1.448	&	-0.677	&	0.000	&	-2.255	&	0.289	\\
(e)	&	0.000	&	-7.507	&	0.000	&	0.000	&	-3.439	&	-0.952	&	0.000	&	-4.195	&	0.292	\\
(f)	&	0.000	&	-4.341	&	0.000	&	0.011	&	-1.038	&	-0.611	&	0.002	&	-1.642	&	0.286	\\
(g)	&	0.000	&	-7.447	&	0.000	&	0.000	&	-4.108	&	-0.605	&	0.000	&	-4.742	&	0.296	\\
(h)	&	0.023	&	-1.110	&	0.000	&	0.159	&	-0.040	&	0.319	&	0.000	&	-9.237	&	0.112	\\
(i)	&	0.004	&	-1.577	&	0.000	&	0.024	&	-0.884	&	-0.407	&	0.000	&	-16.106	&	0.102	\\
(j)	&	0.019	&	-1.130	&	0.000	&	0.334	&	0.794	&	1.210	&	0.000	&	-3.579	&	0.128	\\
(k)	&	0.000	&	-3.916	&	0.000	&	0.002	&	-1.632	&	-0.016	&	0.000	&	-3.818	&	0.141	\\
\end{tabular}
\end{table}

\section{Proofs and Derivations}
\subsection{Proof of Proposition \ref{prop:Tbar}}
For each $i$,
\begin{align*}
E({T}^{(\tau_0)}_{\Gamma i} \mid \cF, \cZ) &= \sum_{i=1}^{n_i}\varrho_{ij}(\delta_{ij} - \tau_0 - \vartheta_{\Gamma ij})\\ &= \sum_{j=1}^{n_i}\varrho_{ij}\left\{\delta_{ij} - \tau_0 - \sum_{k=1}^{n_i}\varrho^{(j)}_{ik}(\delta^{(j)}_{ik}-\tau_0)\right\}
\\&\leq \sum_{j=1}^{n_i}\varrho_{ij}\left\{\delta_{ij} - \tau_0 - \sum_{k=1}^{n_i}\varrho_{ik}(\delta^{(j)}_{ik}-\tau_0)\right\}\\
&= \sum_{j=1}^{n_i}\varrho_{ij}\left\{(1-\varrho_{ij})(\tau_{ij} - \tau_0)+ (1-\varrho_{ij})(\tau_{ij}-\tau_0)/(n_i-1)\right\}\\
&= \frac{n_i}{n_i-1}\sum_{j=1}^{n_i}\varrho_{ij}(1-\varrho_{ij})(\tau_{ij}-\tau_0).
\end{align*}
Taking the weighted average of these terms yields the result.

\subsection{Proof of Theorem \ref{thm:impossible}}
 Without loss of generality, assume $h_{\Gamma n_i}(\hat{\tau}_i-\tau_0)$ passes through the origin; otherwise simply define $\tilde{h}_{\Gamma n_i}(\hat{\tau}_i - \tau_0) = h_{\Gamma n_i}(\hat{\tau}_i - \tau_0)  - h_{\Gamma n_i}(0)$. We first consider the case $n_i=3$, and suppress the potential dependence of $h_{\Gamma n_i}$ on $n_i$ in what follows. Suppose there are $B=3$ matched sets, with potential outcomes and hidden covariates
of the following form

\begin{center}
\begin{tabular}{c c c c}
& $r_T$ & $r_C$ & $u$\\
{Set 1} & $a+2c$ & $a$ & 1\\  
&$a$ & $a$&0\\
&$a$ &$a$&0\\
\hline
{Set 2} & $a-c$ & $a$&0\\
&$a$ & $a$&1\\
&$a$ &$a$&1\\
\hline
{Set 3} & $a-c$ & $a$&0\\
&$a$ & $a$&1\\
&$a$ &$a$&1\\
\end{tabular}
\end{center}
Assume $c>0$, and observe $\bar{\tau} = 0$. It is clear that if the second or third individual in any set receives the treatment, the observed treated-minus-control paired difference will be zero, and the worst-case expectation generated under the assumption of Fisher's sharp null will also be zero. We focus attention on what occurs if the first individual receives the treatment. If this occurs in set one, then $\hat{\tau}_1=\delta^{(1)}_{11} = c$. Further, assuming Fisher's sharp null, the unobserved treated-minus-control paired differences would be improperly imputed as $\delta^{(1)}_{12} = \delta^{(1)}_{13} = -c$, while in reality they are $\delta_{12} = \delta_{13} = 0$. In sets two and three, if the first individual receives the treatment then $\hat{\tau}_i=\delta_{i1} = -c$, and under the assumption of the sharp null the worst-case expectation would be calculated assuming the unobserved treated-minus-control paired differences were $\delta^{(1)}_{i2}=\delta^{(1)}_{i3} = c/2$, while in reality $\delta_{i2} = \delta_{i3} = 0$ for $i=2,3$.

In the first set, if the first individual received the treatment, the worst-case confounder under Fisher's sharp null would assign the treatment to individual 1 with probability $\Gamma/(2+\Gamma)$, and to individuals 2 and 3 with probability $1/(2+\Gamma)$ each, which align with the true probabilities of assignment to treatment under the true vector $\bu_1$. For matched set 1, the expectation of $h_{\Gamma}(\hat{\tau}_i) - \mu_{\Gamma i}$ is
\begin{align*}
&\frac{\Gamma}{2+\Gamma}\left\{h_{\Gamma}(2c) - \left(\frac{\Gamma h_{\Gamma}(2c)}{2+\Gamma} + \frac{h_{\Gamma}(-c)}{2+\Gamma} + \frac{h_{\Gamma}(-c)}{2+\Gamma}\right)\right\}\\
&= \frac{\Gamma}{(2+\Gamma)^2}\left\{2h_{\Gamma}(2c) - 2h_{\Gamma}(-c)\right\}
\end{align*}
For sets two and three, if the first individual receives the treatment then the worst-case confounder under Fisher's sharp null would instead assign the treatment to individual 1 with probability $1/(2\Gamma + 1)$, and to individuals 2 and 3 with probability $\Gamma/(2\Gamma+1)$. These probabilities also align with the true assignment probabilities under $\bu_2$ and $\bu_3$. The expectation of $h_{\Gamma}(\hat{\tau}_i) - \mu_{\Gamma i}$ for either of these sets would be
\begin{align*}
&\frac{1}{2\Gamma+1}\left\{h_{\Gamma}(-c) - \left(\frac{ h_{\Gamma}(-c)}{2\Gamma+1} + \frac{\Gamma h_{\Gamma}(c/2)}{2\Gamma+1} + \frac{\Gamma h_{\Gamma}(c/2)}{2\Gamma+1}\right)\right\}\\
&= \frac{\Gamma}{(2\Gamma+1)^2}\left\{2h_{\Gamma}(-c) - 2h_{\Gamma}(c/2)\right\}.
\end{align*}

The sum of the expectations across the three matched sets is
\begin{align*}
\frac{\Gamma}{(2+\Gamma)^2}\left\{2h_{\Gamma}(2c) - 2h_{\Gamma}(-c)\right\} +\frac{2\Gamma}{(2\Gamma+1)^2}\left\{2h_{\Gamma}(-c) - 2h_{\Gamma}(c/2)\right\},\end{align*}
and for the expectation to be positive for some $\Gamma$, it suffices to show that there exists a $\Gamma$ such that
\begin{align*} \left\{2h_{\Gamma}(2c) - 2h_{\Gamma}(-c)\right\} > \left(\frac{2+\Gamma}{2\Gamma+1}\right)^2\left\{-4h_{\Gamma}(-c) + 4h_{\Gamma}(c/2)\right\}\end{align*}
Take $\Gamma = 1 + 3/\sqrt{2} + \epsilon$ for some $\epsilon > 0$, which in turn implies $(2+\Gamma)^2/(2\Gamma+1)^2 = 1/2 - 1/4 \iota_\epsilon$ for some $0 < \iota_\epsilon < 1$. Recalling that $h_{\Gamma}$ passes through the origin and is non-decreasing, we have $h_\Gamma(-c) \leq 0\leq h_\Gamma(c/2)\leq h_\Gamma(2c)$. We then have 
\begin{align*}&\left\{2h_{\Gamma}(2c) - 2h_{\Gamma}(-c)\right\} + (2+\Gamma)^2/(2\Gamma+1)^2\left\{4h_{\Gamma}(-c) - 4h_{\Gamma}(c/2)\right\}\\
&= \left\{2h_{\Gamma}(2c) - 2h_{\Gamma}(-c)\right\} + (1/2 - 1/4\iota_\epsilon)\left\{4h_{\Gamma}(-c) - 4h_{\Gamma}(c/2)\right\}\\
&= 2h_{\Gamma}(2c) - \iota_\epsilon 2h_{\Gamma}(-c) + (2 - \iota_\epsilon)h_{\Gamma}(c/2) \\
&\geq  - \iota_\epsilon 2h_{\Gamma}(-c) + (4 - \iota_\epsilon)h_{\Gamma}(c/2).
\end{align*}
By assumption $h_{\Gamma}(\cdot)$ is nondecreasing and not constant. Therefore, there exists a $c$ such that $h_{\Gamma}(-c) < h_{\Gamma}(c/2)$ Choosing this value of $c$, we then have 
$- \iota_\epsilon 2h_{\Gamma}(-c) + (4 - \iota_\epsilon)h_{\Gamma}(c/2) > 0$, proving the result.

The construction naturally extends to other values of $n_i$. Choose $n_i$ strata, each of size $n_i$, where the first has one individual with $r_{Tij} = a + (n_i-1)c, r_{Cij} = a$ and the rest have $r_{Tij}=r_{Cij}=a$. Let the remaining $n_i-1$ strata each have one individual with $r_{Tij} = a -c, r_{Cij} = a$ and the rest $r_{Tij}=r_{Cij}=a$. 
\subsection{Proof of Proposition \ref{prop:interval}}
Observe that $\sum_{j=1}^{n_i}(\delta_{ij}-\tau_0) = n_i(\bar{\tau}_i - \tau_0)$. Define $\tilde{\delta}_{ij}$ as 
\begin{align}\label{eq:pairex}
\tilde{\delta}_{ij} &= \begin{cases} {2(\delta_{ij} - \tau_0)}/{(1+\Gamma)} & \delta_{ij} -\tau_0\geq 0\\
 {2\Gamma (\delta_{ij}-\tau_0)}/{(1+\Gamma)}& \delta_{ij} - \tau_0 < 0,\end{cases} 
\end{align}
such that ${D}^{(\tau_0)}_{\Gamma i} = \sum_{i=1}^{n_i}Z_{ij}\tilde{\delta}_{ij}$ and $E({D}^{(\tau_0)}_{\Gamma i}\mid \cF, \cZ) = \sum_{i=1}^{n_i}\varrho_{ij}\tilde{\delta}_{ij}$. By ignoring the constraint that $\sum_{i=1}^{n_i}\varrho_{ij} = 1$, it is clear that $E({D}^{(\tau_0)}_{\Gamma i}\mid \cF, \cZ)$ is upper bounded under (\ref{eq:interval}) in the manuscript by setting
\begin{align*}
\varrho_{ij} &= \begin{cases}   \Gamma/\kappa_{\Gamma i} & \tilde{\delta}_{ij} \geq 0\\
 1/\kappa_{\Gamma i} & \tilde{\delta}_{ij} < 0.\end{cases} 
\end{align*}
Doing so yields 
\begin{align*}
E({D}^{(\tau_0)}_{\Gamma i}\mid \cF, \cZ) &= \sum_{j=1}^{n_i}\varrho_{ij}\left\{\delta_{ij} - \tau_0 - \left(\frac{\Gamma-1}{1+\Gamma}\right)|\delta_{ij} - \tau_0|\right\}\\
&\leq \left(\frac{2\Gamma}{1+\Gamma}\right)\frac{\sum_{j=1}^{n_i}(\delta_{ij} - \tau_0)}{\kappa_{\Gamma i}}= n_i\left(\frac{2\Gamma}{1+\Gamma}\right)\frac{(\bar{\tau}_i-\tau_0)}{\kappa_{\Gamma i}}
\end{align*}

\subsection{Proof of Proposition \ref{prop:equiv}}
 We can write $W^{(\tau_0)}_{\Gamma i}$ in the form $\underset{\kappa_{\Gamma i}^{-1}\leq \varrho_{ij} \leq \Gamma\kappa^{-1}_{\Gamma i}}{\min} \sum_{i=1}^{n_i}Z_{ij}(\delta_{ij} - \tau_0)/(n_i\varrho_{ij})$. For each $ij$, we have \begin{align}\label{eq:exipw}
\underset{\kappa_{\Gamma i}^{-1}\leq \varrho_{ij} \leq \Gamma\kappa^{-1}_{\Gamma i}}{\min}\;\; \frac{(\delta_{ij} - \tau_0)}{n_i\varrho_{ij}} &= \begin{cases} \kappa_{\Gamma i}{(\delta_{ij} - \tau_0)}/(n_i\Gamma) & \delta_{ij} -\tau_0\geq 0\\
 \kappa_{\Gamma i}{(\delta_{ij} - \tau_0)/(n_i)} & \delta_{ij} - \tau_0 < 0.\end{cases} 
\end{align}
Comparing (\ref{eq:exipw}) to (\ref{eq:pairex}), we see that 
\begin{align*}
W^{(\tau_0)}_{\Gamma i}  &= \frac{\kappa_{\Gamma i}}{n_i}\left(\frac{1+\Gamma}{2\Gamma}\right){D}^{(\tau_0)}_{\Gamma i}.
\end{align*}
\subsection{Proof of Proposition \ref{prop:zero}}
For any vector of unmeasured confounders $\bu_i$,
\begin{align}\label{eq:numAi}
E_{\bu_i}(A_i\mid \cF, \cZ) &= \frac{\sum_{j=1}^{n_i}\exp(\gamma u_{ij})[q_{ij} - \left\{(\Gamma-1)/(1+\Gamma)\right\}|q_{ij}|]}{\sum_{j=1}^{n_i}\exp(\gamma u_{ij})}
\end{align}
When $q_{ij} \geq 0$,  $q_{ij} - \left\{(\Gamma-1)/(1+\Gamma)\right\}|q_{ij}|$ equals $2q_{ij}/(1+\Gamma)$, and it equals $2\Gamma q_{ij}/(1+\Gamma)$ for $q_{ij} < 0$. It is clear that the numerator in (\ref{eq:numAi}) is maximized by setting $u_{ij} =  \1(q_{ij} \geq 0)$. Under this choice, the numerator becomes $2\Gamma/(1+\Gamma)\sum_{j=1}^{n_i}q_{ij} = 0$  as $\sum_{j=1}^{n_i}q_{ij} = 0$.  The choice $u_{ij} =  \1(q_{ij} \geq 0)$ thus also maximizes (\ref{eq:numAi})
\subsection{Proof of Theorem \ref{thm:bound}}

Suppose (\ref{eq:sensmodel}) holds at $\Gamma$, and let $\bu_i$ be a vector of unmeasured confounders. We first show that
\begin{align*} E_{\bu_i}({D}^{(\tau_0)}_{\Gamma i}\mid \cF, \cZ) &\leq E_{\bu_i}({D}^{(\bar{\tau}_i)}_{\Gamma i}\mid \cF, \cZ) + \left(\frac{2\Gamma}{1+\Gamma}\right)\left\{1 + \frac{1-\Gamma}{\Gamma}\P_{\bu_i}(\hat{\tau}_i \geq \bar{\tau}_i\mid \cF, \cZ)\right\}(\bar{\tau}_i-\tau_0). 
\end{align*}Let $\varrho_{ij} = \exp(\gamma u_{ij})/\sum_{j=1}^{n_i}\exp(\gamma u_{ij})$. The function 
\begin{align*}
f_i(t) &:= \sum_{i=1}^{n_i}\varrho_{ij}\left\{\delta_{ij} - t  - \left(\frac{\Gamma-1}{1+\Gamma}\right)|\delta_{ij} - t|\right\}
\end{align*}
is concave in $t$, and is differentiable everywhere except the points $\{\delta_{ij}, j=1,...,n_i\}$. Observe that $f_i(t) = E_{\bu_i}({D}^{(t)}_{\Gamma i} \mid \cF, \cZ)$. By concavity, we have that for any superderivative $v$ of $f_i$ at $t = \bar{\tau}_i$,

\begin{align*}
E_{\bu_i}({D}^{(\tau_0)}_{\Gamma i} \mid \cF, \cZ) \leq E_{\bu_i}({D}^{(\bar{\tau}_i)}_{\Gamma i} \mid \cF, \cZ) + v(\tau_0 - \bar{\tau}_i).
\end{align*}
One superderivative $v'$ of $f(t)$ at $f(\bar{\tau}_i)$ is 
\begin{align*}
v' &= -\frac{2\Gamma}{1+\Gamma}\sum_{j=1}^{n_i}\varrho_{ij}\left\{\1(\delta_{ij} < \bar{\tau}_i) + \1(\delta_{ij} \geq \bar{\tau}_i)/\Gamma
\right\}\\
&= -\frac{2\Gamma}{1+\Gamma}\left(1 + \frac{1-\Gamma}{\Gamma}\sum_{j=1}^{n_i}\varrho_{ij}\1(\delta_{ij}\geq \bar{\tau}_i)\right)
\\
&= -\frac{2\Gamma}{1+\Gamma}\left(1 + \frac{1-\Gamma}{\Gamma}\P_{\bu_i}(\hat{\tau}_i \geq \bar{\tau}_i\mid \cF, \cZ)\right),
\end{align*}
and using this superderivative gives the desired inequality.

To see that $E_{\bu_i}({D}^{(\bar{\tau}_i)}_{\Gamma i}\mid \cF, \cZ) \leq 0$ if (\ref{eq:sensmodel}) holds at $\Gamma$, note that $\sum_{i=1}^{n_i}(\delta_{ij} - \bar{\tau}_i)= 0$, and that ${D}^{(\bar{\tau}_i)}_{\Gamma i} = \sum_{j=1}^{n_i}Z_{ij}(\delta_{ij}-\bar{\tau}_i - \left\{(\Gamma-1)/(1+\Gamma)\right\}|\delta_{ij}-\bar{\tau}_i|)$. By Proposition \ref{prop:zero}, we then have that $f_i(\bar{\tau}_i) = E_{\bu_i}({D}^{(\bar{\tau}_i)}_{\Gamma i} \mid \cF, \cZ)\leq 0$ if (\ref{eq:sensmodel}) holds at $\Gamma$. The suitably weighted average of these terms proves the theorem.
\section{On regularity conditions for Proposition \ref{prop:conservative} and Theorems \ref{thm:inference2} - \ref{thm:binary}}
\
Proposition \ref{prop:conservative} states that under suitable regularity conditions, then conditional upon $\cF$ and $\cZ$
\begin{align*}
\underset{B\rightarrow\infty}{plim}\;\; \frac{\text{se}_\bQ^2(\bar{D}^{(\tau_0)}_\Gamma)}{\text{\var}(\bar{D}^{(\tau_0)}_\Gamma\mid \cF, \cZ)}&\geq 1.
\end{align*}
This is required to justify replacing the true, but unknowable, value for $\text{\var}(\bar{D}^{(\tau_0)}_\Gamma\mid \cF, \cZ)$ with the conservative standard error $\text{se}_\bQ^2(\bar{D}^{(\tau_0)}_\Gamma)$ when performing inference. For the result to hold, it is sufficient to show that $B\times \text{se}_\bQ^2(\bar{D}^{(\tau_0)}_\Gamma)$ converges in probability to a limiting value as $B\rightarrow \infty$; and that $B\times \text{\var}(\bar{D}^{(\tau_0)}_\Gamma\mid \cF, \cZ)$ converges to a limiting value as $B\rightarrow \infty$. For finely stratified experiments, Theorem 2 of \citet{fog18mit} provides a proof of these details, with Conditions 1-3 therein providing sufficient conditions. Suitable modification of these regularity conditions also yields the result in a sensitivity analysis with $\Gamma>1$. The analogous derivation would follow through with $\bar{D}^{(\tau_0)}_\Gamma$ replaced with $\tilde{K}^{(\tau_0)}_\Gamma$

For Theorems \ref{thm:inference2} or \ref{thm:binary} to hold, armed with Proposition \ref{prop:conservative} we need only show that a central limit theorem holds when applied to the true, but unknowable, deviate
\begin{align*}
\frac{\bar{D}^{(\tau_0)}_{\Gamma}  - E_{\bu}(\bar{D}^{(\tau_0)}_\Gamma \mid \cF, \cZ)}{\var_\bu(\bar{D}^{(\tau_0)}_\Gamma \mid \cF, \cZ)^{1/2}}.
\end{align*}
For Theorem \ref{thm:inference1} to hold, an analogous argument shows we simply need a central limit theorem to hold with $\bar{D}^{(\tau_0)}_\Gamma$ replaced with $\tilde{K}^{(\tau_0)}_\Gamma$.

To see why, suppose that (\ref{eq:sensmodel}) holds at $\Gamma$ and that $\bar{\tau} = \tau_0$, let $\bu$ be the true (but unknowable) vector of unmeasured confounders, and observe that for $\bar{D}^{(\tau_0)}_{\Gamma} \geq 0$,
\begin{align}\label{eq:idealdeviate}
\frac{\bar{D}^{(\tau_0)}_{\Gamma}  - E_{\bu}(\bar{D}^{(\tau_0)}_\Gamma \mid \cF, \cZ)}{\var_{\bu}(\bar{D}^{(\tau_0)}_\Gamma \mid \cF, \cZ)^{1/2}}  &\geq \frac{\bar{D}^{(\tau_0)}_{\Gamma}  - \max_{\bu\in \mathcal{U}}\;E_{\bu}(\bar{D}^{(\tau_0)}_\Gamma \mid \cF, \cZ)}{\var_{\bu}(\bar{D}^{(\tau_0)}_\Gamma \mid \cF, \cZ)^{1/2}}\nonumber\\& \gtrapprox \frac{\bar{D}^{(\tau_0)}_{\Gamma}  - \max_{\bu\in \mathcal{U}}\;E_{\bu}(\bar{D}^{(\tau_0)}_\Gamma \mid \cF, \cZ)}{\text{se}(\bar{D}^{(\tau_0)}_\Gamma)}. 
\end{align} 
The first line is obvious. The second line would hold as a deterministic inequality when $\bar{D}^{(\tau_0)}_\Gamma \geq 0$ with $\text{se}(\bar{D}^{(\tau_0)}_\Gamma)$ replaced by $E\{\text{se}^2(\bar{D}^{(\tau_0)}_\Gamma)\mid \cF, \cZ\}^{1/2}$, and Proposition \ref{prop:conservative} implies that the standard error, suitably scaled, converges to this expectation in probability under suitable regularity conditions. If the standardized deviate on the left-hand side of (\ref{eq:idealdeviate}) obeys a central limit theorem, applying Slutsky's lemma using the results of Proposition \ref{prop:conservative} then yields that for any $k > 0$
 \begin{align*}
\lim_{B\rightarrow \infty} \P\left(\frac{\bar{D}^{(\tau_0)}_{\Gamma}  - \max_{\bu\in \mathcal{U}}\;E_{\bu}(\bar{D}^{(\tau_0)}_\Gamma \mid \cF, \cZ)}{\text{se}(\bar{D}^{(\tau_0)}_\Gamma)} \geq k \right) &\leq \lim_{B\rightarrow \infty}\P\left(\frac{\bar{D}^{(\tau_0)}_{\Gamma}  - E_{\bu}(\bar{D}^{(\tau_0)}_\Gamma \mid \cF, \cZ)}{\var_{\bu}(\bar{D}^{(\tau_0)}_\Gamma \mid \cF, \cZ)^{1/2}} \geq k\right) \\&= 1-\Phi(k),
\end{align*} 
such that inference using the deviate (\ref{eq:idealdeviate}) and employing a standard normal reference distribution would facilitate an asymptotically valid sensitivity analysis at $\Gamma$ for $H_N^{(\tau_0)}$. \\

Sufficient conditions for a central limit theorem would include, for instance, satisfying the Lyapunov condition: for some $\delta > 0$,
\begin{align*}
\underset{B\rightarrow\infty}{\lim}\frac{\sum_{b=1}^B(n_i/N)^{2+\delta}E_{\bu_i}\{|{D}_{\Gamma_i}^{(\tau_0)} - E_{\bu_i}({D}_{\Gamma_i}^{(\tau_0)}\mid \cF, \cZ)|^{2+\delta}\mid \cF, \cZ\}}{\var_\bu(\bar{D}^{(\tau_0)}_{\Gamma} \mid \cF, \cZ)^{1+\delta/2}} = 0
\end{align*}
Many reasonable restrictions on $\cF$ would satisfy this condition. For finely stratified experiments, Theorem 1 of \citet{fog18mit} provides a proof that central limit theorem applies, with Conditions 1-3 therein providing sufficient conditions.

\section{A reference distribution built from biased randomizations}
In paired observational studies, \citet{fog19student} develops an alternative reference distribution for conducting a sensitivity analysis. Rather than using $\Phi(\cdot)$, one can instead consider constructing a reference distribution based upon the randomization distribution for $\bar{D}_\Gamma^{(\tau_0)}/\text{se}_\bQ(\bar{D}_\Gamma^{(\tau_0)})$ under $H_F^{(\tau_0)}$ while using the vector of unmeasured confounders yielding the worst-case expectation when assuming constant effects at $\tau_0$. Explicitly, for a sensitivity analysis assuming (\ref{eq:sensmodel}) holds at $\Gamma$, consider constructing a reference distribution $\hat{G}_\Gamma(\cdot)$ as follows
\begin{enumerate}
\item For each $ij$, compute $a_{ij} = (R_{ij} - Z_{ij}\tau_0) - \sum_{j'\neq j}(R_{ij'} - Z_{ij'}\tau_0)/(n_i-1)$
\item Set $u_{ij} = \1(a_{ij}\geq 0)$
\item Define $\hat{G}_\Gamma(\cdot)$ as 
\begin{align*}
\hat{G}_\Gamma(k) &= \sum_{\bz\in \Omega}\1\left\{\frac{\bar{A}_\Gamma(\bz, \ba)}{\text{se}_\bQ\{\bar{A}_\Gamma(\bz, \ba)\}}\leq k\right\}\prod_{i=1}^B \frac{\exp\left(\gamma \sum_{j=1}^{n_i}z_{ij}u_{ij} \right)}{\sum_{j=1}^{n_i}\exp(\gamma u_{ij})},
\end{align*}
where $A_{\Gamma i} = \sum_{j=1}^{n_i}z_{ij}[a_{ij} - \{(\Gamma-1)/(1+\Gamma)\}|a_{ij}|]$, $\bar{A}_\Gamma(\bz, \ba) = \sum_{i=1}^B(n_i/N)A_{\Gamma i}$, and $\text{se}_\bQ\{\bar{A}_\Gamma(\bz, \ba)\}$ is the corresponding standard error based on $(n_i/N)A_{\Gamma i}$, $i=1,...,B$.
\end{enumerate}
Observe that under $H_N^{(\tau_0)}$, $\hat{G}_\Gamma(\cdot)$ is random as $a_{ij}\neq r_{Cij} - \sum_{j'\neq j}r_{Cij'}/(n_i-1)$. Consider replacing $\hat{\varphi}^{(\tau_0)}(\alpha,\Gamma)$ with 
\begin{align*}
\hat{\varphi}^{(\tau_0)}_{rand}(\alpha,\Gamma) &= \1\left\{\frac{\bar{D}^{(\tau_0)}_\Gamma}{\text{se}_\bQ(\bar{D}^{(\tau_0)}_\Gamma)}\geq \hat{G}_{\Gamma}^{-1}(1-\alpha)\right\}\end{align*}

where $\hat{G}_{\Gamma}^{-1}(1-\alpha) = \inf\{k: \hat{G}_\Gamma(k)\geq  1-\alpha\}$  is the $1-\alpha$ quantile of $\hat{G}_\Gamma(\cdot)$.

\begin{proposition}\label{prop:rand}
Under suitable regularity conditions, for all points $k$ and conditional upon $\cZ$ and $\cF$.
\begin{align*}
\hat{G}_{\Gamma}(k)\overset{p}{\rightarrow}\Phi(k).
\end{align*}
\end{proposition}

The flow of the proof is identical to that of Theorem 2 in \citet{fog19student}, and a sketch is as follows. One considers the joint distribution of $\sqrt{B}\bar{A}_\Gamma(\bz, \ba)$ and $\sqrt{B}\bar{A}(\bz', \ba)$, for $\bz$, $\bz'$ $iid$ from worst-case distribution for treatment assignments constructed above. One then shows that $\{\sqrt{B}\bar{A}_\Gamma(\bz, \ba), \sqrt{B}\bar{A}(\bz', \ba)\}$ are identically distributed and converge jointly in distribution to a multivariate normal, with mean zero and covariance zero, such that they are asymptotically independent. One further shows that $B\times \text{se}_\bQ(\bar{A}_\Gamma(\bz, \ba))^2$ is consistent for the variance of $\sqrt{B}\bar{A}_\Gamma(\bz, \ba)$, such that $\bar{A}(\bz, \ba)/\text{se}_\bQ(\bar{A}(\bz, \ba))$ converges in distribution to a standard normal. Theorem 15.2.3 of \citet{leh05} then completes the proof.

Proposition \ref{prop:rand} implies that $\hat{\varphi}^{(\tau_0)}(\alpha,\Gamma)$ may be replaced with $\hat{\varphi}^{(\tau_0)}_{rand}(\alpha,\Gamma)$ in the statement of Theorem \ref{thm:inference2}. At $\Gamma=1$, we further have that if $H_F^{(\tau_0)}$ is true, $E\{\hat{\varphi}_{rand}^{(\tau_0)}(\alpha,1)\mid \cF, \cZ\} \leq \alpha$ for all $B>p$, where $p$ was the dimension of the matrix $\bQ$ used to construct the standard error. That is, in finely stratified experiments,  $\hat{\varphi}_{rand}^{(\tau_0)}(\alpha,1)$ is exact under constant effects and asymptotically correct under $H_N^{(\tau_0)}$, thus extending the results of \citet{loh17} and \citet{wu18} to this experimental design. Unfortunately, when (\ref{eq:sensmodel}) holds at $\Gamma>1$ it need not be the case that $E\{\hat{\varphi}_{rand}^{(\tau_0)}(\alpha,\Gamma)\mid \cF, \cZ\}\leq \alpha$ in finite samples under $H_F^{(\tau_0)}$. This is because in using the pattern of unmeasured confounding that maximizes the worst-case expectation in constructing $\hat{\varphi}_{rand}^{(\tau_0)}(\alpha,\Gamma)$, we are already appealing to an asymptotic argument provided in \citet{gas00} and outlined in \S \ref{sec:cons} for how to approximate the worst-case $p$-value: in finite samples, simply maximizing the worst-case expectation does not necessarily yield the worst-case $p$-value. When conducting sensitivity analyses with matched structures beyond pair matching, one generally cannot efficiently calculate the true worst-case $p$-value even under constant effects. Nonetheless, one may expect $\hat{\varphi}_{rand}^{(\tau_0)}(\alpha,\Gamma)$ to better capture finite-sample departures from normality in the distribution of $\bar{D}_\Gamma^{(\tau_0)}/\text{se}_\bQ(\bar{D}_\Gamma^{(\tau_0)})$.

\bibliographystyle{apalike}
\bibliography{bibliography.bib}

\begin{thebibliography}{}

\bibitem[Aronow and Lee, 2012]{aro12}
Aronow, P.~M. and Lee, D.~K. (2012).
\newblock Interval estimation of population means under unknown but bounded
  probabilities of sample selection.
\newblock {\em Biometrika}, 100(1):235--240.

\bibitem[Bai et~al., 2019]{bai19}
Bai, Y., Shaikh, A., and Romano, J.~P. (2019).
\newblock Inference in experiments with matched pairs.
\newblock Technical Report 2019-63, University of Chicago, Becker Friedman
  Institute for Economics.

\bibitem[Chung and Romano, 2013]{chu13}
Chung, E. and Romano, J.~P. (2013).
\newblock Exact and asymptotically robust permutation tests.
\newblock {\em The Annals of Statistics}, 41(2):484--507.

\bibitem[Chung and Romano, 2016]{chu16}
Chung, E. and Romano, J.~P. (2016).
\newblock Multivariate and multiple permutation tests.
\newblock {\em Journal of Econometrics}, 193(1):76--91.

\bibitem[Ding, 2017]{din17}
Ding, P. (2017).
\newblock A paradox from randomization-based causal inference.
\newblock {\em Statistical Science}, 32(3):331--345.

\bibitem[Fogarty, 2018]{fog18mit}
Fogarty, C.~B. (2018).
\newblock On mitigating the analytical limitations of finely stratified
  experiments.
\newblock {\em Journal of the Royal Statistical Society: Series B (Statistical
  Methodology)}, 80:1035--1056.

\bibitem[Fogarty, 2019]{fog19student}
Fogarty, C.~B. (2019).
\newblock Studentized sensitivity analysis for the sample average treatment
  effect in paired observational studies.
\newblock {\em Journal of the American Statistical Association}, DOI:
  10.1080/01621459.2019.1632072.

\bibitem[Fogarty et~al., 2017]{fog17}
Fogarty, C.~B., Shi, P., Mikkelsen, M.~E., and Small, D.~S. (2017).
\newblock Randomization inference and sensitivity analysis for composite null
  hypotheses with binary outcomes in matched observational studies.
\newblock {\em Journal of the American Statistical Association},
  112(517):321--331.

\bibitem[Gastwirth et~al., 2000]{gas00}
Gastwirth, J.~L., Krieger, A.~M., and Rosenbaum, P.~R. (2000).
\newblock Asymptotic separability in sensitivity analysis.
\newblock {\em Journal of the Royal Statistical Society: Series B (Statistical
  Methodology)}, 62(3):545--555.

\bibitem[Hansen, 2004]{han04}
Hansen, B.~B. (2004).
\newblock Full matching in an observational study of coaching for the {SAT}.
\newblock {\em Journal of the American Statistical Association},
  99(467):609--618.

\bibitem[Horvitz and Thompson, 1952]{hor52}
Horvitz, D.~G. and Thompson, D.~J. (1952).
\newblock A generalization of sampling without replacement from a finite
  universe.
\newblock {\em Journal of the American Statistical Association},
  47(260):663--685.

\bibitem[Janssen, 1997]{jan97}
Janssen, A. (1997).
\newblock Studentized permutation tests for non-iid hypotheses and the
  generalized {Behrens-Fisher} problem.
\newblock {\em Statistics and Probability Letters}, 36(1):9--21.

\bibitem[Kang et~al., 2016]{kan16}
Kang, H., Kreuels, B., May, J., and Small, D.~S. (2016).
\newblock Full matching approach to instrumental variables estimation with
  application to the effect of malaria on stunting.
\newblock {\em The Annals of Applied Statistics}, 10(1):335--364.

\bibitem[Lehmann and Romano, 2005]{leh05}
Lehmann, E.~L. and Romano, J.~P. (2005).
\newblock {\em Testing statistical hypotheses}.
\newblock Springer Science \& Business Media.

\bibitem[Loh et~al., 2017]{loh17}
Loh, W.~W., Richardson, T.~S., and Robins, J.~M. (2017).
\newblock An apparent paradox explained.
\newblock {\em Statistical {S}cience}, 32(3):356--361.

\bibitem[Miratrix et~al., 2017]{mir17}
Miratrix, L.~W., Wager, S., and Zubizarreta, J.~R. (2017).
\newblock Shape-constrained partial identification of a population mean under
  unknown probabilities of sample selection.
\newblock {\em Biometrika}, 105(1):103--114.

\bibitem[Neyman, 1935]{ney35}
Neyman, J. (1935).
\newblock Statistical problems in agricultural experimentation.
\newblock {\em Supplement to the Journal of the Royal Statistical Society},
  2(2):107--180.

\bibitem[Pashley and Miratrix, 2019]{pas19}
Pashley, N.~E. and Miratrix, L.~W. (2019).
\newblock Insights on variance estimation for blocked and matched pairs
  designs.
\newblock {\em arXiv preprint arXiv:1710.10342}.

\bibitem[Rosenbaum, 1991]{ros91}
Rosenbaum, P.~R. (1991).
\newblock A characterization of optimal designs for observational studies.
\newblock {\em Journal of the Royal Statistical Society. Series B
  (Methodological)}, 53(3):597--610.

\bibitem[Rosenbaum, 1995]{ros95}
Rosenbaum, P.~R. (1995).
\newblock Quantiles in nonrandom samples and observational studies.
\newblock {\em Journal of the American Statistical Association},
  90(432):1424--1431.

\bibitem[Rosenbaum, 2002]{obs}
Rosenbaum, P.~R. (2002).
\newblock {\em {Observational Studies}}.
\newblock Springer, New York.

\bibitem[Rosenbaum, 2013]{ros13}
Rosenbaum, P.~R. (2013).
\newblock Impact of multiple matched controls on design sensitivity in
  observational studies.
\newblock {\em Biometrics}, 69(1):118--127.

\bibitem[Rosenbaum, 2020]{ros20}
Rosenbaum, P.~R. (2020).
\newblock Modern algorithms for matching in observational studies.
\newblock {\em Annual Review of Statistics and Its Application}, 7(1).

\bibitem[Rosenbaum and Krieger, 1990]{ros90}
Rosenbaum, P.~R. and Krieger, A.~M. (1990).
\newblock Sensitivity of two-sample permutation inferences in observational
  studies.
\newblock {\em Journal of the American Statistical Association},
  85(410):493--498.

\bibitem[Sabbaghi and Rubin, 2014]{sab14}
Sabbaghi, A. and Rubin, D.~B. (2014).
\newblock Comments on the {Neyman-Fisher} controversy and its consequences.
\newblock {\em Statistical Science}, 29(2):267--284.

\bibitem[Stuart, 2010]{stu10}
Stuart, E.~A. (2010).
\newblock Matching methods for causal inference: A review and a look forward.
\newblock {\em Statistical Science}, 25(1):1--21.

\bibitem[Stuart and Green, 2008]{stu08}
Stuart, E.~A. and Green, K.~M. (2008).
\newblock Using full matching to estimate causal effects in nonexperimental
  studies: examining the relationship between adolescent marijuana use and
  adult outcomes.
\newblock {\em Developmental Psychology}, 44(2):395--406.

\bibitem[Wu and Ding, 2018]{wu18}
Wu, J. and Ding, P. (2018).
\newblock Randomization tests for weak null hypotheses.
\newblock {\em arXiv preprint arXiv:1809.07419}.

\bibitem[Zhao et~al., 2019]{zha19}
Zhao, Q., Small, D.~S., and Bhattacharya, B.~B. (2019).
\newblock Sensitivity analysis for inverse probability weighting estimators via
  the percentile bootstrap.
\newblock {\em Journal of the Royal Statistical Society: Series B (Statistical
  Methodology)}, DOI: 10.1111/rssb.12327.

\end{thebibliography}
\end{document}